\documentclass[11pt]{article}

\usepackage[]{amsmath,amssymb}
\usepackage{amsfonts}
\usepackage{graphics,epsfig}

\usepackage{amsfonts}
\let\sect=\section
\def\section{\newpage\sect}

\def\text#1{\mbox{\rm #1\ }}

\def\ie{{\rm i.e.,\/}\ }
\def\etc{{\rm etc.\/}\ }

\newcommand{\RR}{\mathbb{R}}

\newcommand{\HH}{\mathbb{H}}
\title{
Large scale geometry and evolution of a universe with radiation pressure and
cosmological constant.
   \vspace{0.8cm}}

\author{R. Coquereaux${}^{1,}$ ${}^2$ \thanks{~Email:
Robert.Coquereaux@cpt.univ-mrs.fr}$\;$,
        A. Grossmann${}^1$\thanks{~Email: Alex.Grossmann@genetique.uvsq.fr} \\
\\
${}^1$ {\it Centre de Physique Th\'eorique - CNRS} \\
       {\it Campus de Luminy - Case 907}           \\
       {\it F-13288 Marseille - France}            \\
\\
${}^2$ {\it CERN} \\
       {\it CH-1211 - Gen\`eve 23}\\
       {\it Switzerland}\\
}

\date{}
\begin{document}
\thispagestyle{empty}
\begin{titlepage}
\maketitle
\abstract{In view of new experimental results that strongly suggest a 
non-zero cosmological constant, it becomes interesting to revisit the
Friedman-Lema\^\i tre model of evolution of a universe with 
cosmological constant and radiation pressure. In this paper, we 
discuss the explicit solutions for that model, and perform numerical 
explorations for reasonable values of cosmological parameters.
We also analyse the behaviour of redshifts in such models 
and the description of  ``very large scale geometrical features'' 
when analysed by distant observers. }

\vspace{4. cm}

\noindent Subject headings: \\
\noindent cosmology:theory---cosmological parameters--- 
 large-scale structure of universe



\vspace{2. cm}

\noindent {\tt  astro-ph/0101369}\\
\noindent CERN-TH/2000-180\\
\noindent CPT-2000/P.4109 \\

\vspace*{0.3 cm}

\end{titlepage}



\section{INTRODUCTION} 
 
According to present experimental constraints, the universe, nowadays, 
is almost spatially flat; moreover, the effect of the radiation term, in the 
Friedman's equation, is also negligeble. However, one should distinguish 
between  the 
description of the universe, as  it appears ``now'',and the description 
the description 
of how it was and how it will be. 
 
Most papers dealing with the determination of the cosmological 
parameters, nowadays, deal with the universe as it appears 
``experimentally'', now. 
Many other articles deal with the complicated issues related with the 
mechanism(s) of inflation, reheating \etc

In the present article, starting with recent experimental constraints 
(\cite{Perlmut},\cite{PerlmutBNL},\cite{CosmicTri},\cite{Efstathiou}) 
on the density parameters $\Omega_{m}$, $\Omega_{\Lambda}$, 
we want to study analytically  the whole history of the 
universe, \ie how it was ({\it after} the first few minutes) and how it 
wil be. A detailed analytic discussion of Friedman universes with 
cosmological constant and radiation pressure was performed years 
ago (\cite{CoqueGros}, \cite{DabroStel}), but at that time, 
the cosmological constant was believed to be zero by many, so that 
the above work -- although justified by a particular phenomenological 
analysis ( \cite{Souriau}) -- could be considered only  as only a purely 
mathematical exercise.

At the light of recent experimental results, part of the work carried 
out in (\cite{CoqueGros}, \cite{DabroStel}) becomes phenomenologically 
relevant : we do not plan to give a thorough discussion of analytical 
solutions in all possible cases, since this was done already by us, 
but restrict our attention to those cases that are compatible with 
recent observations and present an updated analytic discussion. 
 
We are not concerned, in 
this paper, with the problem of inflation (there are several theoretical 
scenarios\ldots) as it is clear that, at such early periods, Friedman's 
equation is anyway not expected to provide a good mathematical model of the 
evolution of the universe. It looks however reasonable to suppose that 
there was a time, in some remote past, 
where the universe was already almost homogeneous and still rather hot. 
One can then use Friedman's equation to extrapolate, back in time, to this 
early epoch. Putting artificially the radiation term to zero is an 
unnecessary simplification that would prevent one to 
perform useful extrapolation towards this remote past. 
One can  actually keep that term at no cost since cosmological models 
with a non-zero 
cosmological constant and radiation pressure can be studied 
analytically. Moreover, the experimental constraints on the radiation 
term are quite well known thanks to the measurement of the Cosmic Microwave 
Background 
radiation (CMB). 
 
A similar comment can be made about the value of the 
constant parameter 
($k=\pm 1$ or $0$) that allows one to distinguish between spatially 
closed ($k=1$) and  open ($k=0$ or $-1$) universes: it may be that the 
reduced cosmological curvature 
density $\Omega_{k}$ is very small, {\it 
nowadays}, and experimentally  compatible with 
zero; however, this quantity is function 
of  time, and  setting artifically the constant $k$ to $0$ 
prevents one from studying some possibly interesting features in 
the history of the universe (for instance, the 
existence of an inflexion point at some particular time during the 
expansion, requires 
$k=1$). 
 
 For the above reasons we shall not drop, in general, the radiation term and 
will keep $k$ as an unknown discrete parameter with values $\pm 1$ 
or $0$. Our main purpose, however, is to study the effects of a non zero (and positive) 
cosmological constant on the analytic solutions describing the dynamics of our  
universe and on the large scale geometry. 
 
In the coming section, we shall recall the relations 
between different cosmological quantities of interest. One should 
remark that experimentally available parameters do not necessary 
coincide with the quantities that provide a good mathematical 
description for the solutions of Friedman's equation.

 In the following section we shall start with this equation (written with a 
``good'' set of variables), and 
 shall analyse possible histories of our universe,  both qualitatively (using 
a well known associated mechanical model \cite{Wheeler}) and analytically (using 
mathematical techniques (that can be traced back to \cite{Lemaitre}) 
 explained in \cite{CoqueGros} and used in 
\cite{CoqueGros}, 
\cite{DabroStel}). 
 
Finally, in the last section, we  investigate the influence of the selection of a 
 cosmological model matters, on ``changes of 
redshift charts''. 
 We answer the following question (following again 
\cite{CoqueGros}): 
if some observer, located somewhere in the universe, 
plots the redshift of every celestial object against its direction, 
what is the redshift that would be measured, for the same objects,  by 
another observer, at the same moment of time (we have an homogeneous 
cosmology!) but situated somewhere else -- and possibly very far -- in 
the spatial universe ? 
These transformation laws involve not only the geometry of our universe 
(relative  location of the two points) but also its dynamics (in  
particular the value of the cosmological constant).

If the distribution of matter in the universe posesses some particular 
large scale features (for instance a ``symmetry'' of some kind), our  
solution of the above problem becomes phenomenologically relevant. 
Indeed, there is no reason to believe that our position (our galaxy) 
is ideally located with respect to such putative elements of symmetry,  
and it would therefore be necessary to perform a change of position to  
recognize such features. For instance, the absence of matter in a  
three dimensional shell centered around some particular point would not 
appear as such when analysed from a galaxy situated far away from this  
special point (this was actually the hypothesis made in \cite{Souriau}).

\bigskip

This paper presents two very different  results: one is concerned with 
analytical solutions describing the dynamics of the universe, the other
is concerned with large scale (and non-euclidean) geometry of our 3-dimensional
space manifold.
In both cases we want to draw the attention of the reader on the novel features
presented here.

When the cosmological constant $\Lambda$ is zero, the dynamics of our universe
can be described in terms of elementary trigonometric (circular or hyperbolic)
functions. 
When $\Lambda$ is non zero, the corresponding Friedmann's differential equations,
 cannot be expressed in terms of trigonometric functions. The remark we make in this paper, (and in an earlier one
 (\cite{CoqueGros}))  
is that that this equation may
be handled with a somewhat more general class of periodic functions.

 Different  explicit expressions can be found in papers by  G. Lema\^\i tre (\cite{Lemaitre}),
 but these expressions, involving elliptic integrals, 
 are not very transparent when written as a function of cosmic
time.

 However, a simple trick --namely, writing Friedmann's  differential equation for 
the inverse ``radius of the universe '' in terms of {\sl conformal} time rather than
for the radius in terms of cosmic time--  allows one to find explicit,
 {\sl analytical} solutions, in terms of very basic elliptic functions,
namely, the Weierstrass elliptic function that have been arounds for well
over a century.  
We have not been able to find any description of this simple trick in 
various books or review articles (e.g. \cite{Weinberg}) 
-- even the most recent ones  like
for instance (\cite{Carroll}) --
devoted to the description of a universe with cosmological constant. 
 
 We should maybe stress the fact that the
Weierstrass elliptic  functions are very simple (if not elementary!)
objects, that generalize both circular and hyperbolic trigonometric functions; 
for example, the Weierstrass ${\cal P}$ function is a doubly periodic function
in the complex plane (in the limit when
one of the periods becomes infinite, it can be expressed in terms of usual
trigonometric functions). It also satisfies interesting
relations, such as a duplication formula analogous to the one valid for 
trigonometric functions. This leads to a relation between cosmological
quantities measured at different times (more about this in section \ref{sec:duplication}).

Another subject, mostly disconnected from the problem of finding
explicit analytical solutions, is the description  of large scale
geometry of our three-dimensional (spatial) universe; this
aspect of our work was  already presented in the above introduction.
However, in order to dissipate possible misunderstandings,
let us stress again the fact that the corresponding geometry
involves the curvature of a $3$-dimensional curved manifold (not like the 
$2$-dimensional sphere which is the object of spherical trigonometry), 
 so that even if our formulae (concerning e.g. arc-length of the sides of a
geodesic triangle) are reminiscent of usual spherical trigonometry,
they are actually quite different. This part of our work 
can be considered as a physically relevant but not totally
trivial mathematical exercise in riemannian geometry.
As a matter of fact, the
corresponding formulae do not seem to be available in mathematical treaties
of differential geometry (probably because they are ``nothing else'' than
a particular example of calculations that one can do on any riemannian 
manifold when its underlying space has the topology of a Lie group).
 
From the (astro) physical point of view, we do not believe that such
formulae have been presented elsewhere.

\section{RELATIONS BETWEEN COSMOLOGICAL PARAMETERS}

\subsection{Geometrical Background for 
Friedman-Lema\^ \i tre Models} 
Most of the present section is well known and can be 
found in standard textbooks. 
The dimensionless 
temperature $T(\tau)$ which seems  preferable to 
the radius $R(t)$) if one wants to perform an analytic 
discussion of Friedman's equation, was introduced by \cite{CoqueGros}. 
 Our purpose here is mostly to provide a 
 summary of useful formulae 
and present our notations. 

\subsubsection{Units} 
We chall work in a system of units in which both $\hbar$ and $c$ 
are set to $1$. All quantities with dimention are thus given as 
powers of length [cm]. 
 
 In our units, energies are in $[cm^{-1}]$. Since  temperatures (to be 
denoted by $\tilde T$ in this paper) appear 
only in products $k_B \tilde T$ (where $k_B$ is the Boltzmann constant), we can take take 
the temperature as dimensionless, and $k_B$ as an energy.

\subsubsection{Metric and Topological Considerations}
As it is well known, these models assume that the universe is, in
first approximation, homogeneous and isotropic.
\paragraph{The metric}
There are three possibilities:
\begin{itemize}
	\item $k=1$,  $ds^{2} = -dt^2 + R^2 [d\chi^2 + \sin^2 \chi(d\theta^2 + \sin^2\theta d\phi^2)]$

	\item $k=-1$,  $ds^{2} =  -dt^2 + R^2 [d\chi^2 + \sinh^2 \chi(d\theta^2 + \sin^2\theta d\phi^2)]$ 

	\item $k=0$,  $ds^{2} =  -dt^2 + R^2 [d\chi^2 +  \chi^2 (d\theta^2 + \sin^2\theta d\phi^2)]$
\end{itemize}

\paragraph{The topology} 
 
The first case $k=1$, is sometimes called the ``compact case'' (closed 
universe),  since the spatial universe is then compact; the global 
topology for space-time can be taken as 
$S^{3}\times \RR$ but one should remember that the above expression 
for the metric is only local, so that one could take, in place of 
$S^{3}$, any quotient $S^{3}/\Gamma$ of the $3$-sphere by a finite 
group $\Gamma$ operating without fixed 
point (for instance one of the binary tetrahedral, octahedral or 
icosahedral subgroups of $SU(2)$).  In 
the case of $S^{3}$, the full isometry group is $SU(2)\times SU(2)$ since 
$S^{3}$ 
itself is homeomorphic with the group of unimodular  and unitary matrices. 
 
In cases $k=0$ or $k=-1$, the space component is non compact (open) and the 
universe is respectively called ``flat'' or ``hyperbolic''. 
 
On intuitive  grounds, and independently of 
experimental fits, one could argue that the model with $k = 1$ is preferable 
to the others since Friedman's equation was derived by assuming 
homogeneity of the stress-energy, something that  is hard to achieve 
in an open universe if the total quantity of matter is finite. 
 
\subsubsection{Friedman's Equation}
This equation is obtained by writing Einstein's equation $G_{\mu\nu} = 8 \pi 
T_{\mu \nu}$ where $T_{\mu \nu}$ is the stress energy tensor of a perfect 
fluid 
and where $G_{\mu\nu}$ is the Einstein tensor associated with the 
above metrics. 
 
The average local energy splits into three parts : 
 \begin{itemize} 
 	\item  Vacuum contribution $\rho_{vac} =\frac{\Lambda}{ 8 \pi G}$ 
 	where ${\Lambda}$ is 
the cosmological constant, 
 
 	\item  Radiation contribution $\rho_{rad}$ such that $R^{4}(t) 
 	\rho_{rad}(t) = const. = \frac{3}{ 8 \pi G} C_{r}$, 
 
 	\item Averaged matter contribution  $\rho_{m}$ such that $R^{3}(t) 
 	\rho_{m}(t) = const. = \frac{3}{ 8 \pi G}C_{m}$ , 
 \end{itemize}

The evolution of $R(t)$ is governed by the Friedman's equation 
$$ 
\frac{1}{ R^{2}}(\frac{d R}{ d t})^{2} = \frac{C_{r}}{  R^{4}} + \frac{C_{m}}{ 
R^{3}} - \frac{k}{ R^{2}} + \frac{\Lambda}{ 3} 
$$ 
It is convenient to introduce the {\sl conformal time} $\tau$, 
defined, up to an additive constant by 
$$ 
d \tau = \frac {d t}{ R} 
$$ 
This variable is natural, both geometrically (it gives three-dimensional 
geodesic distances) and analytically (as we shall 
see later). The quantities $C_{r}$, $C_{m}$ and $\Lambda$ have 
dimensions $L^{2}$, $L$, $L^{-2}$ respectively. In order to replace 
$R$ by a dimensionless quantity, we introduce the characteristic length scale 
of matter $$\Lambda_{c} = \frac{4}{ 9 C_{m}^{2}}$$ and replace $R(\tau)$ 
by the {\sl reduced dimensionless temperature} \cite{CoqueGros} 
\begin{equation} 
 T(\tau) = \frac{1}{ \Lambda_{c}^{1/2} R(\tau)} 
\label{eq:TfromR} 
\end{equation} 
With this new variable, Friedman's equation becomes 
\begin{equation} 
(dT /d\tau)^{2} = \alpha T^{4} + {2\over 3} T^{3} - k T^{2} + {\lambda \over 
3} 
\label{eq:Friedman} 
\end{equation} 
where 
$$ 
\lambda = {\Lambda \over \Lambda_{c}} 
$$ 
and 
$$ 
\alpha =  C_{r} \Lambda_{c} 
$$ 
are two constant dimensionless parameters.

The reason for calling $T$ the ``reduced dimensionless temperature'' is that 
it is proportionnal to the temperature $\tilde T$ of the black body 
radiation. Indeed, $\rho_{rad} = 4 \sigma {\tilde T}^{4}$ where 
$\sigma$ is the Stefan-Boltzmann constant. Since $\rho_{rad} = {3 
\over 8 \pi G} \alpha \Lambda_{c} T^{4}$, one finds 
\begin{equation} 
\tilde T^{4} = {3 \over 8 \pi G} {\alpha \Lambda_{c} \over 4 \sigma} T^{4} 
\label{eq:UsualTemperature} 
\end{equation} 
 
This  may be the right place to recall that Lema\^\i tre (\cite{Lemaitre}) 
did, long ago, an analytic study of the solutions of Friedman's equations; 
his discussion, made in terms of $R$ and $t$ involves, of course, elliptic functions, 
but it is only when we express the cosmological quantities 
in terms of  $T(\tau)$ (the reduced temperature, as a function 
of conformal time) that these quantities can be written themselves as 
elliptic functions with respect to a particular lattice (the 
associated Weierstrass ${\cal P}$ function appears then naturally).  
The fact that such a direct link with the theory of Weierstrass 
can be established should be clear from the fact 
that, with our parametrization, the right hand side of equation \ref{eq:Friedman} 
is a polynomial of degree four. 
 
{\sl The Hubble function} decribing the rate of expansion is 
defined, as usual, by  $H = R^{-1}(dR/dt)$ and can be written, in 
terms of $T(\tau)$ as 
$$ 
H(\tau) = - \Lambda_{c}^{1\over 2} {dT \over d\tau} 
$$ 
 
Another useful quantity is the {\sl deceleration function} $ q = - 
R(dR/dt)^{-2} (d^{2}R /dt^{2})$ that can be expressed in terms of 
$T(\tau)$ as follows: 
 
$$ 
q = {- {\lambda \over 3} + {1 \over 3} T^{3} + \alpha T^{4} \over 
 {\lambda \over 3} - k T^{2} + {2 \over 3} T^{3} + \alpha T^{4}} 
$$ 
Actually, the choice of the sign (and of the name) turns out to be a 
historical mistake, since the recent bounds on the cosmological constant 
lead to a negative value 
for $q$ (an accelerating expanding universe). 
 
By multiplying Friedman's equation by $\Lambda_c / H^2$, one obtains the 
famous relation :

\begin{equation} 
1 = \Omega_r + \Omega_m + \Omega_k + \Omega_\Lambda 
\label{eq:CosmicTriangle} 
\end{equation} 
 
with 
\begin{eqnarray} 
\Omega_k &=& - k T^2 {\Lambda_c \over H^2} \\ 
 \Omega_m &=& {2\over 3}  T^3 {\Lambda_c \over H^2} \\ 
\Omega_\Lambda &=& {\lambda \over 3}{\Lambda_c \over H^2}\\ 
 \Omega_r &=& \alpha T^4 {\Lambda_c \over H^2} 
\label{eq:TheOmegas} 
\end{eqnarray} 
 
Notice that the famous ``cosmic triangle'' of \cite{CosmicTri} -- a 
kind of cosmic Dalitz' plot, becomes a triangle only if one decides to forget 
the radiation term (then, indeed, $1 = \Omega_m + \Omega_k + \Omega_\Lambda $). 
In terms of the usual temperature $\tilde T$, the radiation terms reads: 
 
\begin{equation} 
\Omega_r = {32 \pi G \over 3}{\sigma {\tilde T}^4 \over H^2} 
\label{eq:Omegarad} 
\end{equation} 
 
\subsubsection{Description of the History of the Universe by the Reduced 
Temperature Function $T(\tau)$} 
 
\begin{itemize} 
    \item Rather than using $R(t)$, we describe \cite{CoqueGros} the history of the 
    universe by the function $T(\tau)$. This function can be 
    intuitively thought of as a dimensionless inverse radius;
it is proportionnal to the usual temperature $\tilde T$.
    Because of the phenomenon 
    of expansion, these quantities $T$ and $\tilde T$ decrease with $\tau$. The argument 
    $\tau$ itself, a dimensionless arc length, measures time : 
    for instance, if the spatial universe is 
    (hyper)spherical,  $\tau$ is a measure (in radians) of the length of 
the arc 
    spanned by a photon that was produced at the big bang. A given 
    solution (a given ``history'' of a -- dimensionless -- universe) is fully 
    characterized by the solution of a differential equation 
    depending on the two dimensionless parameters $\alpha$ and 
    $\lambda$. Of course, one has also to specify some initial value 
    data (one takes $T(\tau) \rightarrow \infty$ when $\tau \rightarrow 
    0$). As we shall see in a later section, $T(\tau)$ is a particular 
elliptic 
    function; the history of our universe is then also fully encoded by the two 
Weierstrass invariants ($g_2, g_3$), or, equivalently, by the two periods 
$\omega_1,\omega_2$ of this elliptic 
function. The Weierstrass invariants are given, in terms of 
parameters $k$, $\alpha$ and $\lambda$, by the algebraic relations 
 
\begin{equation} 
g_{2} = {k^{2} \over 12} + {\alpha \lambda \over 3} 
\label{eq:WeierstrassInvariantsg2}. 
\end{equation} 
 
\begin{equation} 
g_{3} = {1 \over 6^{3}} (k^{3} - 2 \lambda) - {{\alpha \lambda k} \over 
18 } 
\label{eq:WeierstrassInvariantsg3}. 
\end{equation} 
The formulae giving the two periods $\omega_1,\omega_2$ in terms of the 
cosmological 
parameters $k$, $\alpha$ and $\lambda$ (or $g_{1},g_{2}$) involve elliptic 
integrals and 
will be given later. 
 
	\item  Dimensional quantities are obtained from $T(\tau)$ thanks to 
	the measurement of a single function having dimensions of a length. 
	Such a function is usually the Hubble function $H = H(\tau)$. 
	The behaviour of $R(t)$ is given by the two parametric equations 
	$R(\tau)$ and $t(\tau)$. As we shall see later, in all cases of 
	physical interest, $t(\tau)$ reaches a logarithmic singularity for a 
	finite conformal time $\tau_{f}$ (the universe expands forever, 
	but as $t \rightarrow \infty$ the arc length associated 
	with the path of a single photon goes to a finite value $\tau_{f}$). 
 
	\item Finally, it remains to know ``when'' we are, \ie the age of 
	the universe. This can be expressed in terms of the dimensionless 
	quantity $\tau_{o}$ (a particular 
	value of $\tau$) or, more conventionally, in terms of 
	$t_{o}=t(\tau_{o})$. Experimentally, one measures the Hubble 
``constant'' $H_{o} = H(\tau_{o})$, \ie the 
	value of the function $H(\tau)$, now. 
\end{itemize}

\subsubsection{Cosmological Quantities} 
 
We just give here a list of most cosmological quantities of interest.

\paragraph{Dimensionless Quantities}
We have the constant parameters $\alpha$, $\lambda$, $k = 0, \pm 1$,
the time-dependent dimensionless densities $\Omega_k (\tau)$,
$\Omega_\Lambda(\tau)$, $\Omega_m(\tau)$, $\Omega_r(\tau)$ and the reduced
temperature function
$T(\tau)$. Of interest also is the deceleration function $q(\tau)$. Of
course, the conformal time $\tau$ itself  is a non-constant (!)
dimensionless quantity. The black-body temperature $\tilde T$ (temperature of the
	blackbody microwave radiation, experimentally expressed in degree Kelvin ) 
	is itself dimensionless -- see our remark at the beginning of this 
	section. It is proportional to
	the dimensionless reduced temperature $T$ (or to the inverse radius
	$R$).
 
\paragraph{Dimensional quantities} 
\begin{itemize} 
	\item  The Hubble function $H(\tau)$. This quantity is usually chosen 
	to fix the length scale. 
 
	\item  The critical length scale $\Lambda_{c}$. It can be thought of 
	as giving a measure of the ``total mass'' of the universe. More 
	precisely, if the 
	universe is spatially closed and has the topology and metric of a 
	$3$-sphere, its total mass is $M = 4\pi^{2} R^{3} 
	\rho_{m}$ so that $\Lambda_{c} = ({\pi \over 2 G M})^{2}$. 
 
	\item The radius $R$. Intuitivelly, it is a measure of 
	the ``mesh'' of our spatial universe. 
 
	\item The age $t$ of the universe (cosmic time). 
 
\end{itemize}

\subsubsection{Expression of the Cosmological Quantities in Terms of the Hubble Function $H$ and of the Dimensionless Densities 
$\Omega_{\Lambda},\Omega_{k}, \Omega_{m}, \Omega_{r}$ } 
 
Since most experimental results are expressed in terms of the dimensionless 
densities $\Omega_{\Lambda},\Omega_{k}, \Omega_{m}, \Omega_{r}$, we 
express all other cosmological quantities of interest in terms of them. 
Typically, dimensional quantities also involve the value of the 
Hubble function $H(\tau)$. All these formulae can be obtained by 
straightforward algebraic 
manipulations. 
 
\paragraph{ Constant dimensionless quantities (parameters)} 
\begin{equation} 
\lambda = {27\over 4} {\Omega_{\Lambda} \Omega_{m}^{2} \over 
\vert \Omega_{k}^{3} \vert } 
\label{eq:lambda} 
\end{equation} 
 
\begin{equation} 
\alpha = {4\over 9} { \Omega_{r}  \vert \Omega_{k} \vert \over \Omega_{m}^{2}} 
\label{eq:alpha} 
\end{equation} 
 
\paragraph{Time-dependent dimensionless quantities (assuming $k \neq 0$)} 
 
\begin{equation} 
T =  - k {3\over 2} {\Omega_{m} \over  \Omega_{k} } = {3\over 2} 
{\Omega_{m} \over \vert  \Omega_{k} \vert } 
\label{eq:TfromOmega} 
\end{equation} 
 
\begin{equation} 
q = {\Omega_{m} \over 2} - \Omega_{\Lambda} + \Omega_{r} = 
3 {\Omega_{m}\over 2} + {\Omega_{k}} + 2 \Omega_{r} -1 
\label{eq:qfromOmega} 
\end{equation}

The temperature of the CMB radiation:
$$ 
{\tilde T}^{4}  ={3 \over 8 \pi G} {1 \over 4 \sigma} \Omega_{r} H^{2} 
$$

The conformal time $\tau$ itself can be found numerically, once a value of 
$T$ is known, 
by solving the (non algebraic) equation expressing $T$ as a function of 
$\tau$ (see the next section).

\paragraph{Constant dimensional quantities (parameters)} 
\begin{equation} 
\Lambda = 3 H^{2} \Omega_{\Lambda} 
\label{eq:LambdaFromOmega} 
\end{equation} 
 
\begin{equation} 
\Lambda_{c} =  {4 \over 9} {\vert \Omega_{k} \vert^{3} \over \Omega_{m}^{2}} H^{2} 
\label{eq:LambdacFromOmega} 
\end{equation} 
 
$$ 
C_m =  {\Omega_m \over \vert \Omega_{k} \vert^{3/2} H} 
$$ 
 
$$ 
C_r =  {\Omega_r \over \vert \Omega_{k} \vert^{2} H^2} 
$$ 
 
\paragraph{Time-dependent dimensional  quantities  (assuming $k \neq 0$) }

Besides the Hubble ``constant'' $H$ itself, one has to consider also the radius 
\begin{equation} 
R = {1 \over T {\Lambda_{c}}^{1/2}} 
\label{eq:CosmicScale} 
\end{equation} 
the cosmic time (the age of the universe corresponding to conformal  
time $\tau$)  given by 
\begin{equation} 
t(\tau) = {\Lambda_c}^{-1/2} \int_{0}^{\tau} {d \tau' \over T(\tau')} 
\label{eq:CosmicTime} 
\end{equation}

\subsection{Dimensions of the Cosmological Quantities}
 
In our system of units, all quantities are either dimensionless or 
have a dimension which is some power of a lengh ($[cm]$).
We gather all the relevant information as follows:

$$ t \sim R \sim C_{m} \sim [cm] $$

$$ G \sim C_{r} \sim [cm^2] $$

$$ H \sim k_{B} \sim {energy} \sim [cm^{-1}] $$

$$ \Lambda \sim \Lambda_{c} \sim [cm^{-2}] $$

$$ \rho_{vac} \sim \rho_{rad} \sim \rho_{m} \sim \sigma \sim [cm^{-4}] $$

Here $\sigma$ denotes the Stefan constant (remember that degrees Kelvin are 
dimensionless).

Finally we list the dimensionless quantities:

$$\Omega_{m}\sim \Omega_{\Lambda}\sim \Omega_{k}\sim \Omega_{r}\sim q 
\sim T \sim {\tilde T} \sim \tau \sim \alpha \sim \lambda  \sim [cm^{0} = 1] $$

\section{EXPERIMENTAL CONSTRAINTS} 
\subsection{Experimental Constraints from High-redshift Supernovae 
and Cosmic Microwave Background Anisotropies} 
 
In general we shall add an index $o$ to refer to {\sl present} values of 
time-dependent quantities (like $H_o$, $\Omega_k^o$,  $\Omega_r^o$,  $\Omega_m^o$, 
 $\Omega_\Lambda^o$, $q_o$, $T_o$ or $\tilde T_o$). 
 Remember that $k,\alpha, \lambda, \Lambda, \Lambda_c$ are 
constant parameters. 
 
The analysis given by  \cite{Efstathiou} (based on experimental 
results on the latest cosmic microwave background anisotropy and  on 
the distant Type Ia supernovae data \cite{Riess}  \cite{Perlmut}) 
gives  $0.13 <\Omega_{m}^o < 0.43$ and $+ 0.40 < \Omega_{\Lambda}^o < + 0.80$. 
In order to stay on the safe side, we do not restrict ourselves to 
the more severe bounds that one can obtain by performing a combined 
likelihood analysis 
that would use both sets of data coming from anisotropy of CMB radiation 
and distant type Ia supernovae. 
 
Moreover, many experiments, nowadays, give a  
present value of the Hubble function close to $H_{o} = h \, 100 \, km 
\, sec^{-1} 
Mpc^{-1}$, with $h \simeq 0.66$, so that $H_{o}^{-1} \simeq 14.2 \, 
\times 10^{9} \, yr = 13.25 \, \times 10^{27} \, cm$. 
This is the value which we shall use in our estimations. 
 
\subsection{Implications of Experimental Constraints on  
Friedman-Lema\^\i tre Parameters} 
 
With $\hbar = c = 1$, the value of the Stefan-Boltzmann constant is 
$\sigma = {\pi^2 k_B^4 \over 60} = 59.8 cm^{-4}$,
$k_B$ is the Boltzmann constant
 and 
the Newton constant is $G = 2.61 \times 10^{-66} cm^2$. Using the above value for $H_{o}$ 
and formula \ref{eq:Omegarad} (with $\tilde T_o = 2.73 K$), we obtain 
$\Omega_{r}^o \simeq 5.10 \times 10^{-5}$. 
 
Using now the 
``cosmic triangle relation'' (equation \ref{eq:CosmicTriangle}) together 
with the experimental 
bounds on $\Omega_m^o$ and  $\Omega_\Lambda^o$ previously recalled, one finds  
$- 0.23 < \Omega_{k}^o < 0.47 $. 
Remember that, with the present conventions, $\Omega_{k}<0$ when $k = 
+1$. 
 
The formula \ref{eq:LambdaFromOmega} given in the last section leads  to $+ 0.68 \times 10^{-56} \, 
cm^{-2} < \Lambda < + 1.36 \times 10^{-56} \, cm^{-2}$.

In order to obtain good estimates for the remaining quantities, one would 
need a better measurement of the curvature density $\Omega_{k}^o$. Indeed, 
formulae \ref{eq:lambda}, \ref{eq:alpha}, together with the  
previously given bounds on $\Omega_k^o$ only imply the following for the  dimensionless 
constant parameters: 
$\alpha < 6.3 \times 10^{-4}$ and $\lambda > 0.44$.  
 
Notice that $\lambda$ is 
quite sensitive to the 
independent values of  $\Omega_{m}^o$ and $\Omega_{\Lambda}^o$ 
(indeed, $\Omega_k$, appearing in the denominator of equation \ref{eq:lambda}, can be 
very small); for 
instance, the values 
$\Omega_{m}^o = 0.13$, $\Omega_{\Lambda}^o = 0.40$ lead to $\lambda = 0.44$ but 
$\Omega_{m}^o = 0.13$, $\Omega_{\Lambda}^o= 0.80 $ lead to $\lambda = 266.68$. By the way, one should stress the fact that the behaviour of  analytic solutions 
of Friedman's equations is essentially determined by the value of the (dimensionless) 
 reduced cosmological constant $\lambda$, which can be rather ``big'', even when the 
genuine cosmological constant $\Lambda$ (which is a dimensional quantity) is itself very ``small''. 
 
As already mentionned, present experimental constraints only imply $\lambda >  
0.44$, but one should remember that $\Omega_{k} \simeq 1 -  
\Omega_{m} - \Omega_{\Lambda}$ should be negative for a spatially  
closed universe ($k = +1$); therefore,  if, on top of experimental  
constraints on $\Omega_{m}$ and  $\Omega_{\Lambda}$ we {\sl assume}  
that we live in a spatially closed universe (a --reasonable-- hypothesis that we  
shall make later, for illustration purposes) then $\Omega_{m} +  
\Omega_{\Lambda} > 1$ and the constraint on $\lambda$, as given by 
equation \ref{eq:lambda} is more tight; an elementary variational  
calculation shows then that the smallest possible value of $\lambda$,  
taking into account both the experimental constraints and the  
hypothesis $k=+1$, is obtained when both $\Omega_{m}$ and  $\Omega_{\Lambda}$ 
saturate their experimental bounds (values respectively equal to  
$0.43$ and $0.8$), so that $\Omega_{k} \geq - 0.23$ and $\lambda \geq  
82.$ 
 
One should not think that the value of $\lambda $ could be arbitrary large: we 
shall see a little  later (next subsection) that, for experimental  
reasons,  it has also to be 
bounded from above (condition $\lambda_{-} < \lambda < \lambda_{+}$). 
 
Using equations \ref{eq:TfromOmega},  \ref{eq:qfromOmega}, one obtains  
easily 
  the bounds $T_o > 0.41$, $-0.75<q_o<-0.18$ and 
$\Lambda_{c}< 1.6 \times 10^{-56} cm^{-2}$ for the present values of 
the reduced temperature $T$ and  deceleration function $q$ and  
for the critical length scale $\Lambda_c$. 
 
In 
view of the experimental results ($\Omega_{r}^o$ small and  $\Omega_k^o$ 
compatible with zero), one maybe tempted to make the simplifying hypothesis 
$1 = 
\Omega_m^o + \Omega_\Lambda^o$, \ie set {\sl both} terms $\Omega_k^o$ and 
$\Omega_r^o$ to zero in the 
relation $1 = \Omega_m + \Omega_k + \Omega_\Lambda + \Omega_{r}$; 
notice however that  assuming the vanishing of $\Omega_k + \Omega_r$ at all 
times (without assuming 
the vanishing of each term, individually) is totally impossible, as 
these densities are not constant  and it is easy to see, from the 
definition of these quantities, that such a relation can only hold at 
at a single moment; it is obvious that the radiation contribution, 
$\Omega_r^o$, although small, is not 
{\sl strictly\/} zero. Moreover, putting artificially the constant $k$ to zero is 
certainly compatible with the 
present experimental data, but one should be aware of the fact that the 
curvature density $\Omega_k$ is not a constant 
quantity and that setting it to zero at all times is an artificial 
simplification that was probably not justified when the 
universe was younger \ldots

\subsection{A Curious Coincidence}
The parameters $C_{m}$ and $C_{r}$ appearing in Friedman's equation, 
and measuring respectively the  matter contribution and radiation 
contribution to the average local energy, are, {\it a priori \,}, 
independent quantities. For this reason, the relation between $\tilde 
T$ (the temperature of the cosmological background radiation) and 
$T$ (the reduced temperature) involves $C_{m}$ and $C_{r}$.

$$
\tilde T^4 = \frac{3}{8\pi G} \frac{\alpha \Lambda_{c}}{4 \sigma} T^4
=\frac{3}{8\pi G} \frac{C_{r} \frac{2^4}{3^4 C_{m}^4}}{4 \sigma} T^4
$$

One may consider he special case of a universe for which the two 
functions $T(\tau)$ and $\tilde T(\tau)$ coincide; this amounts 
setting $$\frac{C_{r}}{C_{m}^4} = 54 \pi G \sigma = 54 \pi G 
\frac{\pi^2 k_B^4}{60} =  \frac{9}{10} \pi^3  G 
k_B^4$$
In this case, the present values of the densities $\Omega_{m}^o$ and $\Omega_{\Lambda}^o$ 
are no longer independent: $\Omega_{r}^o = 5.1 \times 10^{-5}$ is, as usual, exactly known 
since $\tilde T_{o}= 2.73 K$ is known; again, as usual, we have
$\Omega_{m}^o = \frac{2}{3} \vert \Omega_{k}^o \vert T_{o}$ but if we 
measure $\Omega_{m}^o$ (for instance) and {\sl assume\,} $T_{o} = \tilde T_{o}$,
we find the value of $ \vert \Omega_{k}^o \vert$. Choosing then, for 
instance, $k=+1$ (a closed spatial universe), so that
$\Omega_{k}^o=-\vert \Omega_{k}^o \vert$, we deduce 
$\Omega_{\Lambda}$ from the equation $\Omega_{k}^o + \Omega_{m}^o + 
\Omega_{\Lambda}^o + \Omega_{r}^o = 1$.

We do not see any theoretical reason  that would justify 
to set $T = \tilde T$, since these two quantities could very well be 
different by several orders of magnitude (even with the rather 
small value of the radiation density  $\Omega_{r}^o$),
nevertheless\ldots by a curious 
coincidence, if one takes the two densities $\Omega_{m}^o$ and 
$\Omega_{\Lambda}^o$ to be numerically equal to their highest possible 
values compatible with 
present experimental bounds ($0.43$ and $0.8$), one finds a 
value of $T_o$ equal to $2.80$; this is very near the experimental 
value $2.73$ of the temperature $\tilde T_{o}$. We have no explanation for this 
``cosmological miracle''.

\section{ANALYTIC BEHAVIOUR OF SOLUTIONS} 
 
\subsection{Qualitative Behaviour of Solutions} 
\begin{figure} 
\includegraphics{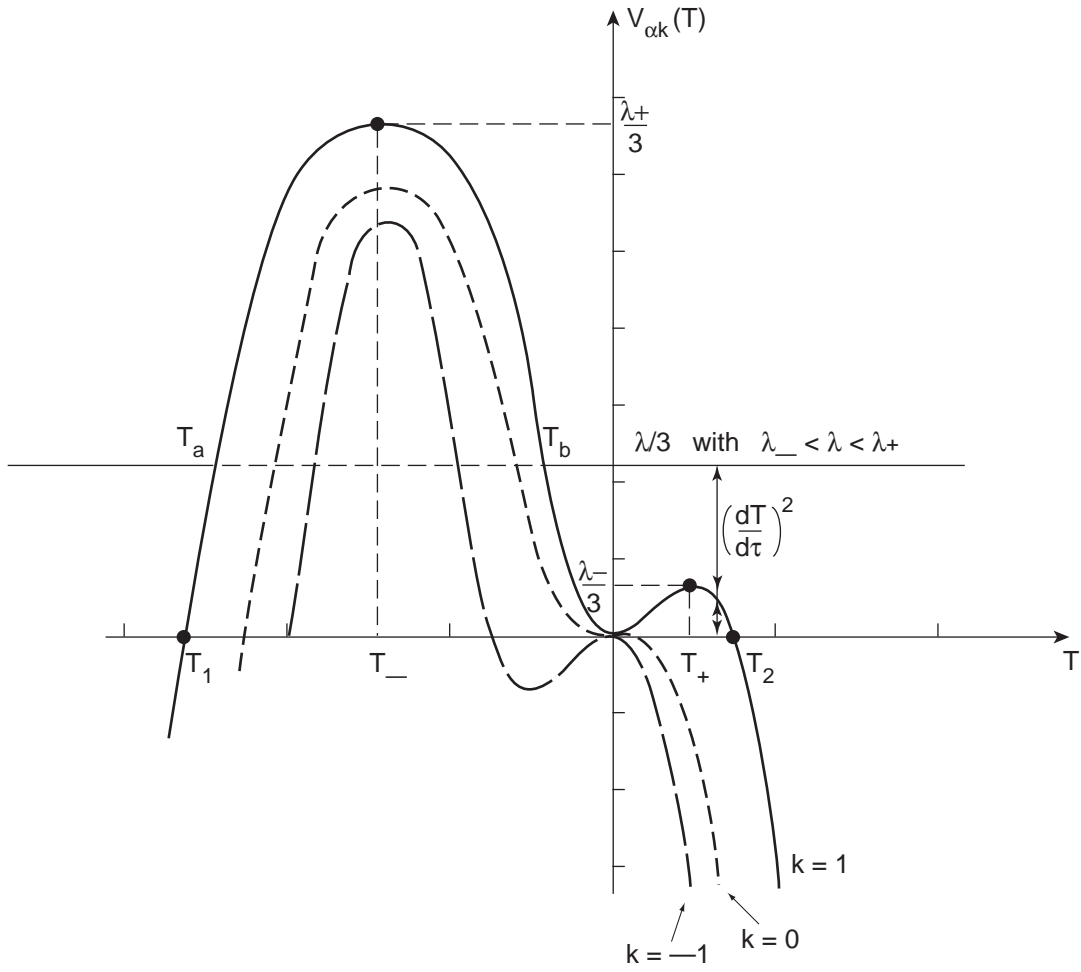} 
\caption{Potential for the associated mechanical system. Case $\alpha  
\neq 0$} 
\label{potential.eps} 
\end{figure}

Equation (\ref{eq:Friedman}) can be written 
$$ 
({dT \over d\tau})^{2} + V_{\alpha, k}(T) = {\lambda \over 3} 
$$ 
with 
$$ 
 V_{\alpha, k}(T) = -\alpha T^{4} -{2\over 3} T^{3} + k T^{2} 
$$ 
This is the equation of a one-dimensional mechanical system with 
``coordinate'' T, potential $V_{\alpha, k}(T)$ (shown in Fig 
\ref{potential.eps}) and total energy $\lambda /3$. The kinetic energy 
beeing non negative, the associated mechanical system describes a 
horizontal line in the $(V(T),T)$ plane but never penetrates under 
the curve $ V_{\alpha, k}(T)$ (that would correspond to $\tau$ 
imaginary). The length of the vertical line segment between a point 
belonging to the curve and  a point with same value of $T$ but 
belonging to the horizontal line $\lambda /3$ (described by the associated 
mechanical 
system) is a measure of $({dT \over d\tau})^{2}$.

For a given value of $\alpha$ (the radiation parameter), and for $k=\pm 
1$, the curve  $V_{\alpha, k}(T)$ 
has typically two bumps (two local maxima). For $k = 1$ (closed 
universe), the right maximum occurs for a positive value of 
$T$ and $V(T)$ whereas, for  $k=-1$ (hyperbolic case), this maximum is 
shifted to $T=0$ 
and $V(T)=0$. For $k=0$ (flat case), the right  maximum 
disappears and we are left with an inflection point at $T=0, V(T)=0$. 
 
\begin{figure} 
\includegraphics*{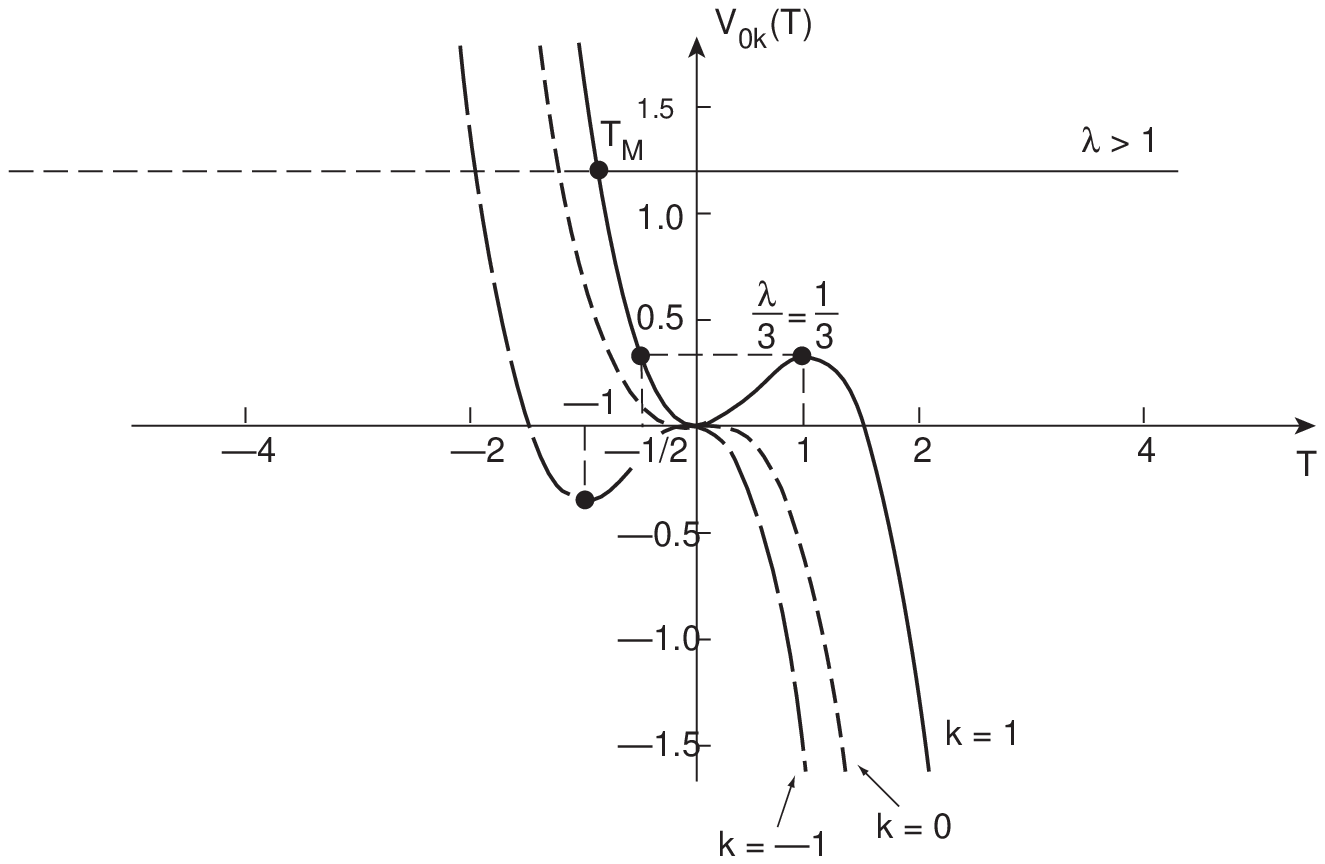} 
\caption{Potential for the associated mechanical system. Case $\alpha =0$} 
\label{potential0.eps} 
\end{figure}

In this paper,we almost always suppose that $\alpha$ is non zero; 
however, one should  notice that if $\alpha = 0$, the curve $V(T)$ 
becomes a cubic (Fig. \ref{potential0.eps}) and the left hand side maximum disappears (it moves 
to $- \infty$ as $\alpha$ goes to $0$). 
 
Notice that figure  \ref{potential.eps} describes only the 
qualitative features 
of our space-time history, since, for reasonable values of $\alpha$ 
and $\lambda$ (\i.e. values compatible with  experimental 
constraints), 
the vertical coordinate of the left hand side maximum should be at least 
$1000$ times higher than the vertical coordinate of the right hand 
side maximum. 
 
The experimental constraints also 
show not only that $\lambda$ is positive (since $\Omega_{\Lambda}$ is) but 
that it is bigger than $\lambda_{-}/3$ (the vertical 
coordinate of the right  maximum) and, at the same time 
 much smaller than $\lambda_{+}/3$ (the vertical 
coordinate of the left  maximum). 
Indeed, for small values of $\alpha$, the positions of the extrema (cf. next 
section) are given by 
$\lambda_{-}\simeq k(1-3 \alpha k)$ and $\lambda_{+}\simeq {1\over 
16\alpha^{3}}(1 + 12 k \alpha)$;  the previous bounds on 
$\lambda$ show then directly that $\lambda > \lambda_{-}$;  
moreover, the 
inequality $\lambda < \lambda_{+}$ is equivalent to $\Omega_{r}^{3} < 
{27\over 256} \Omega_{m}^{4}/\Omega_{\Lambda}$, and this clearly holds. 
Unfortunately, this does not lead to a stringent constraint on 
$\lambda$ istself since the bound $\alpha < 6.3 \times 10^{-4}$ only implies 
 $\lambda_{+} > 2.5 \times 10^8$.

 The associated mechanical system is therefore such that it describes an 
horizontal line like the one 
displayed on Fig. \ref{potential.eps}. 
Typically, a given universe starts from the right of the picture ($T 
\rightarrow 
\infty$ corresponding to the big bang), moves to the left (slowing 
down), until it reaches the vertical axis ($T=0$). This actually takes 
place in 
a finite conformal time $\tau_{f}$ but, as we shall see, it 
corresponds to a cosmic time $t \rightarrow \infty$, so that, for the 
universe in which we live, history stops there. However, the 
solution can be continued for $T<0$ (a negative radius $R$) until the 
system bumps against $V(T)$ to come back to right infinity; the 
system then jumps to left infinity, follows the horizontal line (to the right) till it 
bumps on $V(T)$ again, and comes back. This round trip of the 
associated mechanical system is done in a (conformal) time 
$2\omega_{r}$ -- a period of the corresponding elliptic function. 
Let us stress the fact that only the first part of the motion (from right 
infinity to the intersection with the vertical axis) is physically relevant 
for ordinary matter. 
 
Since the radius $R$ is proportionnal to $1/T$ (see equation \ref{eq:TfromR}), the discussion, in 
terms of $R$ is of 
course different: the system 
starts with $R=0$ (big bang), and 
expands forever; moreover the expansion speeds up in all three cases 
$k=\pm 1$, $k=0$. The only particularity of the case $k = 1$ is that 
there is an inflexion point (coming from the existence of a positive right hand side 
maximum for the curve $V(T)$) : the expansion speeds up anyway, but there is a time $\tau_{I}$ 
for which the rate of expansion vanishes. 
 
\subsection{The Elliptic Curve Associated with a Given Cosmology} 
 
\subsubsection{General Features} 
 
Let us call $Q(T)$ the fourth degree polynomial that appears at the 
right hand side of Friedman's equation (eq. \ref{eq:Friedman}). Let $T_{j}$ be any one of the 
(possibly complex) roots of the equation $Q(T)=0$, then the fractional 
linear transformation 
\begin{equation} 
\label{eq:transformationT} 
 y = {Q'(T) \over 4} {1 \over T-T_{j}} + 
{Q''(T) \over 24} 
\end{equation} brings (\ref{eq:Friedman}) to the form 
\begin{equation} 
\label{WeierstrassDiffEq} 
({ dy \over d\tau})^{2} = P(y) \doteq 4 y^{3} - g_{2} y - g_{3} 
\end{equation} 
where $g_{2}$ and $g_{3}$ are the two Weierstrass invariants given in 
eqs. (\ref{eq:WeierstrassInvariantsg2},\ref{eq:WeierstrassInvariantsg3}).  
The solution to the previous equation is 
well known: $y= {\cal P}(\tau;g_{2},g_{3})$ where ${\cal P}$ is the 
elliptic Weierstrass function associated with the invariants $g_{2}$ 
and $g_{3}$. One should remember that, given a lattice $L$ in the 
complex plane, an elliptic function $f$ with respect to $L$ is non 
constant meromorphic function that is bi-periodic, with respect to 
$L$ (so that $f(\tau + u) = f(\tau)$ whenever $u$ belongs to the 
lattice $L$). It is known (since Liouville) that if $a$ is an arbitrary 
complex number (including infinity), the number of solutions of the 
equation $f(\tau)=a$ is independent of $a$, if multiplicities are 
properly counted; this number is called the {\sl order} of $f$. It is 
useful to know that any rational function of an elliptic function is 
also elliptic (with respect to the same lattice) but that its order 
will coincide with the order of $f$ only if the transformation is 
fractional linear (like the transformation (\ref{eq:transformationT})). 
Finally, one should remember that the order of an elliptic functions is 
at least two and that the Weierstrass ${\cal P}$ function 
corresponding to a given lattice is defined as {\sl the} elliptic 
function of order $2$ that has a pole of order $2$ at the origin (and 
consequently at all other points of $L$) and is such that 
$1/\tau^{2} - {\cal P}(\tau)$ vanishes at $\tau = 0$. The two 
elementary periods generating the lattice $L$ can be expressed in 
terms of $g_{2}$ and $g_{3}$, but conversely, the Weierstrass 
invariants can be expressed in terms of the elementary periods of the 
lattice $L$. 
 
These old theorems of analysis lead directly to the fact that 
$T(\tau)$ is an elliptic function of order $2$ and that other 
quantities that are rational functions of $T$ (like the Hubble 
function $H$, the deceleration function $q$ or all the cosmological 
densities $\Omega_{m}$, $\Omega_{r}$ , $\Omega_{k}$, 
$\Omega_{\Lambda}$) are also elliptic with respect to the same 
lattice (but they are not of order $2$). In this sense, one can say 
that our universe is fully described by the elliptic curve 
$x^{2} = 4 y^{3} - g_{2}y - g_{3}$ characterized by 
the two Weierstrass 
invariants $g_{2}$, $g_{3}$, or, alternatively, by two elementary 
periods generating the lattice $L$. 
 
\begin{figure} 
\includegraphics*{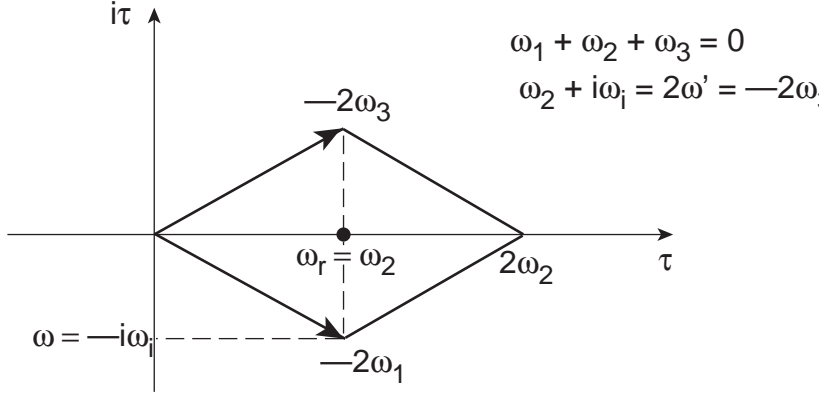} 
\caption{An elementary periodicity cell} 
\label{rhombus.eps} 
\end{figure}

In this (physical) case, the elementary periodicity cell is a rhombus 
(figure \ref{rhombus.eps}) with vertices $\{0,\omega_{r} - i \omega_{i}, 
2 \omega_{r}, \omega_{r} + i \omega_{i}\}$. 
Another standard notations for periods are $2\omega_{2}\doteq 2\omega_{r}$, 
$- 2 \omega_{3} \doteq \omega_{r} + i \omega_{i}$ and $$- 2 \omega_{1} 
\doteq \omega_{r} - i \omega_{i}$$ (so that $\omega_{1} + \omega_{2} + 
\omega_{3} = 0$). The transformation expressing the half-periods 
$\omega_{1,2,3}$ in terms of the Weierstrass invariants $g_{2}$ and 
$g_{3}$ (or conversely) can be obtained, either from a direct numerical 
evaluation of elliptic integrals (see below) 
or from fast algorithms described in \cite{Lautrup};  
one can also 
use the Mathematica function 
$WeierstrassHalfPeriods[\{g_{2},g_{3}\}]$ as well as the 
opposite transformation 
$WeierstrassInvariants[\{\omega_{1},\omega_{2}\}]$. 
In the case $\alpha = 0$, the value $\tau = \omega_r$ turns out to be 
a global minimum of $T(\tau)$, therefore  
another possibility to get $\omega_r$ in that case 
 is to find numerically the first zero  of the 
equation $d T(\tau)/ d \tau = 0$.

\subsubsection{Neglecting Radiation : the Case $\alpha = 0$} 
 \label{sec:duplication}

When $\alpha = 0$, the two poles of $T(\tau)$ coincide, so that the 
dimensionless temperature, up to shift and rescaling, coincides with the 
Weierstrass ${\cal 
P}$-function iself (the fractional linear transformation 
\ref{eq:transformationT} takes the simple form $T = 6 y + k/2$). Notice 
that  ${\cal P}$ is an even function of $\tau$. 
 
\begin{figure} 
\includegraphics*{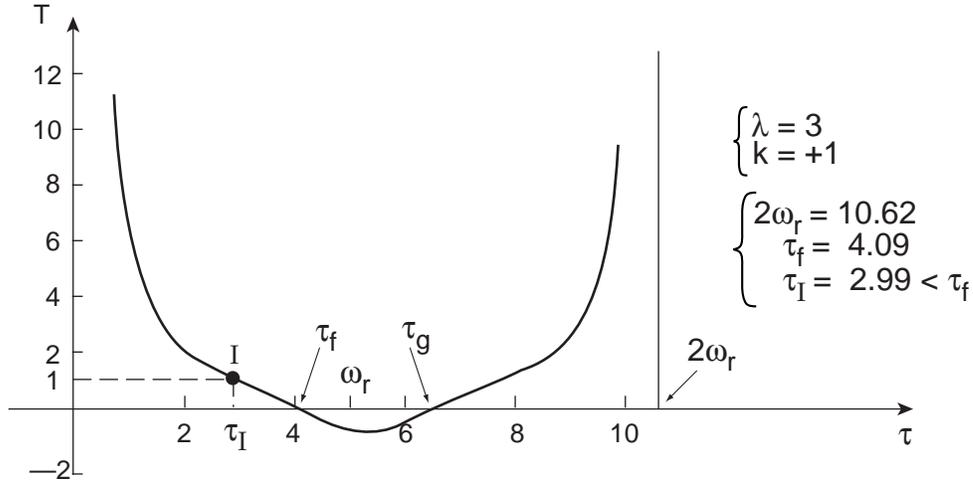} 
\caption{Evolution of the reduced  temperature. Closed case.} 
\label{TgraphClosed.eps} 
\end{figure}

\begin{figure} 
\includegraphics*{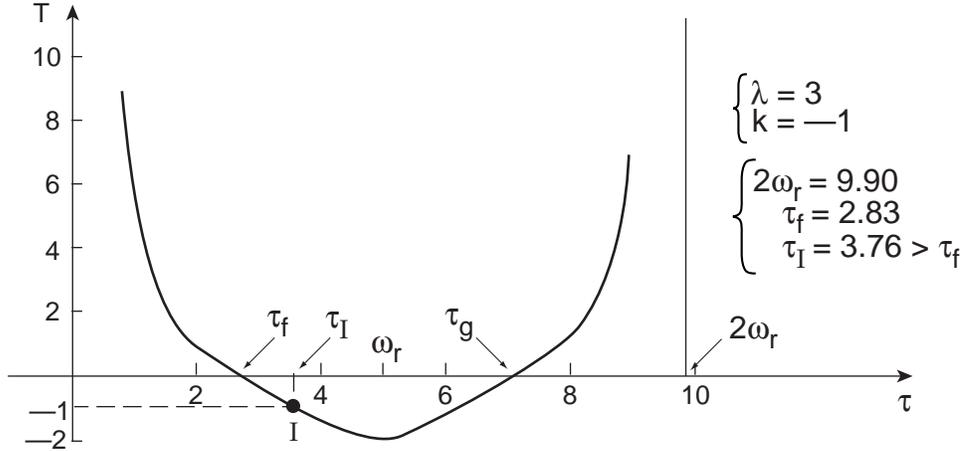} 
\caption{Evolution of the reduced temperature. Open case.} 
\label{TgraphOpen.eps} 
\end{figure}

In that 
case 
\begin{equation} 
T(\tau) = 6[{\cal P}(\tau;g_{2},g_{3}) + k/12] 
\label{eq:TisWeierstrass} 
\end{equation} 
A typical plot of $T(\tau)$ for experimentally reasonable values of 
the cosmological parameters is 
given in figures \ref{TgraphClosed.eps} (closed universe: $k=1$)  
and \ref{TgraphOpen.eps} (open universe: $k=-1$). 
 
In both cases, the physically relevant part of the curve is given by 
the interval $0< \tau < \tau_{f}$ since $\tau = 0$ corresponds to the 
big bang (infinite temperature) and $\tau = \tau_{f}$ to the end of  
conformal time: the cosmic time $t=t(\tau)$ reaches a logarithmic  
singularity when $\tau \rightarrow \tau_{f}$, so that when $t  
\rightarrow \infty$, the arc length described by a photon born with  
the big bang tends toward the finite value $\tau_{f}$; for values of 
$\lambda$ compatible with recent observations, this limit is strictly 
less than $2 \pi$ so that it will never be possible to see the ``back  
of our head'', even if the universe is closed and if 
we wait infinitely many years\ldots. 
 
Notice that both curves show the existence of an inflexion point  
denoted by $\tau_{I}$ on the graphs; however, in the case 
$k=-1$ (open universe), this inflexion point is located after the end 
of conformal time ($\tau > \tau_{f}$) and is therefore physically  
irrelevant. The existence of such a point is of interest only in the case  
of a spatially closed universe. Notice that in that case, the experimental 
observations tell us that the present value $\tau_o$ of $\tau$ (\ie  
todays' date) is bigger than $\tau_{I}$ (and, of course, smaller  
than $\tau_{f}$). 
 
The values $\tau_{f} < \tau < \tau_{g}$ correspond to a classicaly  
forbidden region (negative dimensionless temperature). The values 
$\tau_{g} < \tau < 2 \omega_{r}$ correspond to universe in  
contraction, ending with a big crush\ldots Finally, one should  
remember that $T$ is a (doubly) periodic function, so that the same 
analysis can be performed in all the intervals $[2 p \omega_{r},  
2 (p+1)\omega_{r}]$. 
 
Let us repeat that $\tau$ is a {\sl conformal} time and that 
$t\rightarrow \infty$ when $\tau \rightarrow \tau_{f}$, 
so that, in both cases $k=\pm 1$ we are indeed in a ever expanding 
universe starting with a big bang. The curve obtained for $k=0$ has  
the same qualitative features, but for the fact that the inflection  
point moves to the end of conformal time: $\tau_{I} = \tau_{f}$ in  
that case. 
 
 The shape of $T(\tau)$, as given by 
figures \ref{TgraphClosed.eps}  
and \ref{TgraphOpen.eps} is 
in full agreement 
with the qualitative discussion given in the previous section. Such 
plots 
were already given in \cite{CoqueGros} where 
Weierstrass functions had been numerically calculated from the algorithms 
obtained in the same reference. The same plots can now be obtained 
easily using for instance Mathematica,  (the last 
versions of this program incorporate routines for the  Weierstrass functions ${\cal P}, 
\zeta$ and $\sigma$). Note, however, that intensive calculations involving 
such functions should probably make use of the extremely fast algorithms described in  
reference \cite{Lautrup}; 
these algorithms use, for the numerical evaluation of these functions, a duplication formula 
known for the Weierstrass ${\cal P}$ function. Moreover, evaluation of elliptic integrals (for 
instance caculation of periods in terms of invariants) can also be done by an iterative 
procedure (based on classical properties arithmetic-geometric mean), this is also described 
in reference \cite{Lautrup}. 
 
The duplication formula reads:
$$
{\cal P}(2z) = -2 {\cal P}(z) + \frac{\left[6 {\cal P}^3(z) - g_2/2 \right]^2}
{4\left[4{\cal P}^3(z)- g_2 {\cal P}(z) - g_3\right]}
$$
Therefore, one obtains  a relation
between the values of the reduced temperature $T$ at (conformal) times $\tau$ and  $2\tau$ which,
in the case $k=+1$, reads
$$
T(2\tau) -\frac{1}{2} =
-2(T(\tau) -\frac{1}{2}) + \frac{[(T(\tau)-\frac{1}{2})^3 - \frac{3}{2}]^2}
{4[4(T(\tau)-\frac{1}{2})^3 - 3(T(\tau) - \frac{1}{2}) - (1-2\lambda)]}
$$
When the universe is very young ($\tau$ near zero), one may approximate the ${\cal P}(z)$ function
by $1/z^2$ and therefore the reduced temperature $T(\tau)$ by $6/\tau^2$.

It may be useful to notice, at this point, that when the cosmological constant is zero (so that
we have also $\lambda = 0$ for the reduced cosmological constant), one obtains
$$
T(\tau) = \frac{3}{2 \sin^2 \frac{\tau}{2}}
$$
whereas, when the cosmological constant $\Lambda$ is equal to the critical value $\Lambda_c$ (so that
$\lambda = 1$), one obtains
$$
T(\tau) = 1 + \frac{3}{2 \sinh^2 \frac{\tau}{2}}
$$

As already discussed previously, recent measurements imply that $\lambda >1$, so that
$T(\tau)$ is given by formula \ref{eq:TisWeierstrass}.

\subsubsection{Not Neglecting Radiation : the Case $\alpha \neq 0$} 
 
This case, technically a little bit more difficult,  
is detailed in Appendix 1. 
 
\subsection{Determination of all Cosmological Parameters: a Fictitious Case} 
 
Here we summarize the procedure that should (ideally!) be followed, in order 
to specify all cosmological parameters of interest. 
\subsubsection{A Simple Procedure} 
\begin{itemize} 
\item From experimental results on the  
present values of the Hubble function $H_o$ and of  the 
temperature ${\tilde T}_o$ of the cosmic background radiation,  
 calculate the radiation density $\Omega_r^o$ 
(use formula \ref{eq:Omegarad}). 
 
\item Improve the experimental bounds on the present values of matter density 
$\Omega_m^o$ {\sl and} cosmological constant density $\Omega_\Lambda$ (so, without assuming 
that $k$ is zero, of course). 
 
\item From the ``Cosmic triangle relation'' (eq. \ref{eq:CosmicTriangle}) obtain the 
present value of the curvature density function $\Omega_k^o$. The sign of this quantity 
determines also the  {\sl opposite} of the constant $k = \pm 1$ (or $0$). 
 
	\item  From these results on densities $\Omega$'s, compute the dimensionless 
constant parameters $\alpha$ and $\lambda$ (use equations \ref{eq:alpha} and \ref{eq:lambda}). 
 
	\item  From $\alpha$ and $\lambda$ compute the two Weierstrass 
	invariants $g_{2}$ and $g_{3}$ that characterize the elliptic curve 
associated with our universe (use equations  
\ref{eq:WeierstrassInvariantsg2} and \ref{eq:WeierstrassInvariantsg3}). 
 
\item The present value of $T_o = T(\tau_0)$ 
of the reduced temperature is obtained from equation \ref{eq:TfromOmega}.

	\item  If one decides to neglect radiation ($\alpha \simeq 0$), one can 
	plot directly the reduced temperature in terms of the conformal time 
 $T(\tau) = 6[{\cal P}(\tau;g_{2},g_{3}) + 1/12]$, where 
${\cal P}$ is the Weierstrass elliptic function,  by 
	using, for instance, Mathematica.  
The present value $\tau_o$ of conformal time is obtained by solving numerically 
the equation $T_o =  6[{\cal P}(\tau_o;g_{2},g_{3}) + 1/12]$. 
Other interesting values like $\tau_f$ (the end of conformal time) or $\tau_I$ (inflection 
point, only interesting if $k=+1$) can be determined numerically; for instance $\tau_f$ is 
obtained, with Mathematica, thanks to the function FindRoot (remember that $\tau_f$ is the first 
positive zero of $T(\tau)$). The value of the half-periof $\omega_r$ can be found by evaluation 
of an elliptic integral but it is simpler to determine it   
by solving numerically the equation $\frac{d T(\tau) }{d  
\tau} = 0$ since $\tau = \omega_{r}$ corresponds to a global minimum.  
Another possibility, still with Mathematica, is to use the function 
$WeierstrassHalfPeriods[\{g_2,g_3\}]$ (this function returns a set of two elementary periods, 
and  twice our $ \omega_r$ coincides with the real part of one of them). 
 
	\item  If one decides to keep $\alpha \neq 0$, one has to use the  
	results of Appendix ~1 and 
	determine $\tau_{f}$ and $\tau_{g}$ first; these two real zeros of $T(\tau)$ (within a 
periodicity cell) are given by the elliptic integrals given in Appendix ~1; they have to be 
evaluated numerically (the value $\tau_g$ was just equal to $2\omega_r - \tau_f$ in the case with no 
radiation, but here, it has to be determined separately). The dimensionless temperature $T(\tau)$ 
is then given, in terms of the Weierstrass $\zeta$ function, by formula 
\ref{eq:TfromWeierstrassZeta}.

	\item In all cases, the present value of the deceleration function (a bad name since it 
is experimentally negative!) $q_o$ is given by equation \ref{eq:qfromOmega}. 
 
	\item The value of Hubble constant allows one to determine all the dimensional quantities, in 
particular the critical length parameter $\Lambda_c$ (formula \ref{eq:LambdacFromOmega}),  
the cosmological constant 
$\Lambda$ ($ = \lambda \Lambda_c$), and the present values of the 
the cosmic scale (or ``radius'') $R_o$  (formula \ref{eq:CosmicScale}) 
 and of the cosmic time $t_o$ (formula \ref{eq:CosmicTime}). 
\end{itemize} 
 
\subsubsection{An Example} 
\label{sec:NumericalExample} 
Here we follow the previous procedure assuming fully determined values for the  
Hubble constant and density parameters $\Omega_{m}$ and  
$\Omega_{\Lambda}$. Of course, such precise values are not yet experimentally  
available and we have to make a random choice (compatible with  
observational bounds) in order to illustrate the previous ``simple  
procedure'' leading to the determination of all cosmological  
parameters of interest. In other words, what we are describing here  
is only a possible scenario. 
For simplicity reasons; we shall neglect the  
influence of radiation (so $\alpha \simeq 0$).

\begin{figure}
\includegraphics{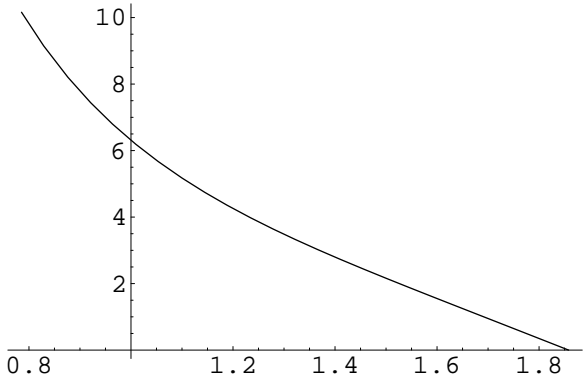}
\caption{Evolution of the reduced temperature in a neighborhood of $\tau_o$}
\label{fig:temperature.eps}
\end{figure}

We take experimental densities $\Omega_{m}=0.4$ and $\Omega_{\Lambda}=0.8$.
Then $\Omega_{k} = -0.2$ and $k=+1$. The reduced cosmological constant
is $\lambda = 108$. The Weierstrass invariants are $g_{2} = 1/12$ and
$g_{3} = -1.00463$. 
The evolution of the dimensionless temperature is $T(\tau) = 6 {\cal 
P}(\tau,\{g_{2},g_{3}\}) + 1/12$ and its present value turns out to be
equal to $T_{o} = 3$; this is a curious coincidence (see section 3.3) 
since $T_{o}$ has no reason, {\it a priori\,}, to be equal to the 
temperature of the cosmological black body radiation $\tilde T_{o}$.
The present value of the conformal time, 
obtained by solving numerically the equation $T(\tau_{o}) = T_o$ is
$\tau_{o} = 1.369$. The end of conformal time, 
obtained by solving numerically the equation $T(\tau_{f}) = 0$ 
is $\tau_{f}=1.858$. The (real) half-period $\omega_{r}$ of the function 
$T(\tau)$ may be obtained by solving numerically the equation
$T(\tau)^\prime = 0$ is $\omega_{r}= 2.662$. The conformal time for 
which $T(\tau)$ has an inflexion point is obtained by solving 
numerically the equation $T(\tau)^{\prime \prime} = 0$ and is 
$\tau_{I} = 1.691$. Notice that in that particular universe, we have
$\tau_{o}< \tau_{I}$ ($< \tau_{f}$), so that we have not 
reached the inflexion point, yet.
Using now the value of the Hubble constant $H_{o} = (13.25 \times 10^{27})^{-1}    
cm^{-1}$, we find a characteristic length scale of matter equal to
$\Lambda_{c} =1.266 \times 10^{-58} cm^{-2}$ so that the cosmological constant itself is equal to
$\Lambda = 1.367 \times 10^{-56} cm^{-2}$.
 Finally we find the cosmic scale
 $R_{o}\doteq R(\tau_{o})= 2.96 \times 10^{28} \,  cm = 3.14 \times 10^{10}\, yrs $ and the 
 cosmic time $t_{o}=t(\tau_{o})=1.24 \times 10^{28} cm$.
Figure \ref{fig:temperature.eps} gives the behaviour of $T(\tau)$ in a neighborhood
of $\tau_o$
 In this example  we neglected the influence of radiation described by the 
 dimensionless parameter $\alpha$ (given by equation \ref{eq:alpha}); using 
 the experimental value of the density parameter $\Omega_{r} = 5.1 \times
 10^{-5}$, one finds $\alpha = 2.77 \times 10^{-5}$. This does not modify
 $T_{o}$ but modifies the values of the Weierstrass invariants $g_{2}$ 
 and $g_{3}$; also, the behaviour of $T(\tau)$ is not the same.
 Taking this value into account together with the results of Appendix 
 1 leads to a slight modification of the values of $\tau_{o}$ and 
 therefore of $t_{o}$.

\section{INFLUENCE OF THE COSMOLOGICAL CONSTANT ON THE 
REDSHIFT. LARGE SCALE STRUCTURES AND GEOMETRY OF THE UNIVERSE.}

\subsection{Cosmological Constant Dependence of the Redshift Function}

\begin{figure}
\includegraphics{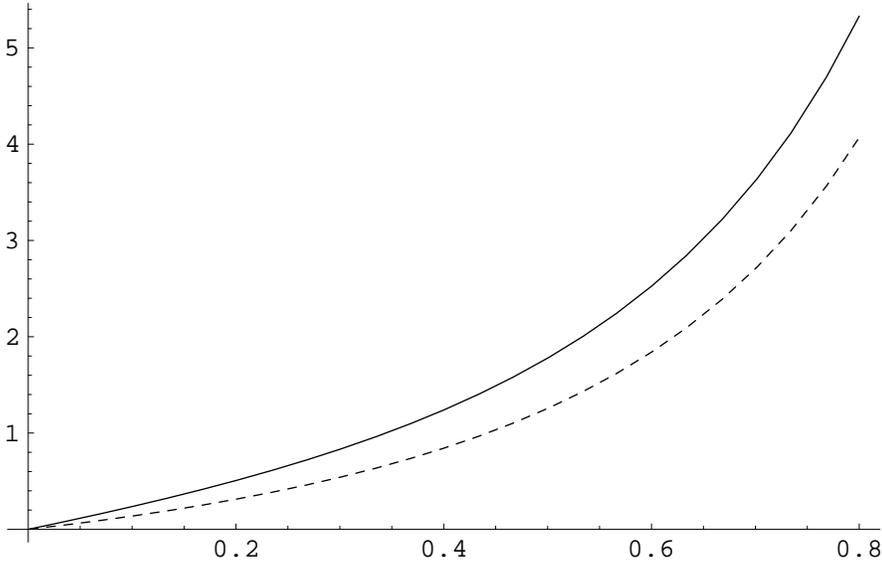}
\caption{Cosmological constant dependence of the redshift function}
\label{fig:redfig.eps}
\end{figure}

We shall now discuss the redshift, $z$, as function of $\tau_{o}$ and 
of the difference 
$$
\delta = \tau_o - \tau
$$
where $\tau$ is parameter time (conformal time) $\tau$ of the emitter and parameter 
time, and $\tau_o$ the parameter time of the observer.
The redshift is given by
$$
z = \frac{R(\tau_o) }{R(\tau)} - 1 = \frac{T(\tau) }{T(\tau_o)} - 1 
$$
Assuming, as before, that we are in the situation $\lambda >1, \alpha =0$,
which is compatible with recent observations,  
we obtain immediately, from equation \ref{eq:TisWeierstrass}, the expression
\begin{equation}
z = C [{\cal P}(\tau_o - \delta) + \frac{1}{12}] - 1
\label{eq:RedshiftFromS}
\end{equation}
with
\begin{equation}
C = [{\cal P}(\tau_o) + \frac{1}{12}]^{-1}
\label{eq:RedshiftConstant}
\end{equation}
Remember that the Weierstrass elliptic function ${\cal P}$
is  characterized by the two Weierstrass invariants $g_{2} =
\frac{1}{12}$  and $g_{3} = \frac{1}{6^3}(k^3 - 2\lambda)$.
In order to illustrate  the influence of the cosmological
constant on the behaviour of the  redshift function, we 
continue our previous example (section \ref{sec:NumericalExample}), therefore taking 
$k=+1$ (a spatially closed universe), $\tau_{o}= 1.3694 $, $\lambda = 
108$ and plot 
$z$ as a function of $\delta$ (see figure \ref{fig:redfig.eps}).
The value $\delta= \tau_o$ (\ie $\tau = 0$)
corresponds to a 
photon that would have been emitted at the Big Bang whereas 
$\delta=0$ (\ie $\tau =\tau_o$) corresponds to a photon emitted by 
the observer himself (no redshift). The dashed curve (the lower curve) gives the 
corresponding redshift when the cosmological constant is absent 
($\lambda=0$).
It is clear from these two curves that the influence of the cosmological 
constant becomes stronger and stronger when the geodesic distance $\delta$ increases. 

\subsection{Observer Dependence of Redshift Values and Large Scale Geometry 
of the universe}

As already announced in the introduction, 
we now want to consider the problem of comparison of ``redshift 
charts'': Given two observers, far apart in the universe, how do we 
compare the redshift that they will record if they look at the same
astronomical object ? 

Actually, one may consider (at least) two very different situations. The 
first is to assume that the two observers perform their measurement at 
the same time (a meaningful notion since we are in a homogeneous 
space-time), in that case, it is clear that the photons that they are 
receiving were not emitted at the same time by the cosmological source
that they are studying.
The other situation is to assume that both are interested in the same 
cosmological event (a star belonging to a distant galaxy turning to a 
supernova, for example) and our two observers record the redshifted 
light coming from this event when the light reaches them (but in 
general this light will not reach them at the same time).

\subsubsection{Comparison of Measurements Made at the Same Time}
We first suppose that both observers perform their observation at 
the same time. It is natural to specify events by a pair $(\sigma, S)$ 
where $S$ is a point in space (Sun for instance) and $\sigma$ a 
particular value of the conformal time of the universe. 
Astronomers on earth ($S$), nowadays ($\tau_o$), may decide to 
observe and record the redshift $z$ of an astronomical object $X$ (a quasar, say).
At the same time $\tau_o$, astronomers located in the 
neighborhood of a star $P$ belonging to a distant galaxy may decide to 
observe the same astronomical object $X$ and record its redshift $Z$.
The problem is to compare $z$ and $Z$. Of course both $z$ and $Z$ are 
given by the previous general formula expressing redshift as a 
function of the time difference between emission and reception, but this value is not the 
same for $S$ and for $P$. In order to  proceed, we need a brief 
discussion involving the geometry of our three-dimensional space 
manifold. 

For definitness, we suppose that we are in the closed case ($k=1$), 
\ie the spatial universe is a three-sphere $S^3$. The reader will have 
no difficulties to generalize our formulae to the open case 
(essentially by replacing trigonometric functions by the 
corresponding hyperbolic ones). The fact that universe expands is 
taken into account by the evolution of the reduced temperature as a 
function of conformal time $\tau$ and, as far as 
geometry is concerned, we may analyse the situation on a fixed 
three-sphere of radius $1$.
The choice of the point $P$ -- we shall call it the ``Pole'' but it is an arbitrary 
point --  allows one to define a notion of equator and of cosmic latitude:
the equator, with respect to $P$, is the two-sphere (a usual sphere) 
of maximal radius, centered on $P$ and the cosmic latitude $\ell(S)$ of the 
Sun $S$ is just the length of the arc of great circle between $S$ and 
the equator; this great circle is the geodesic going through the two 
points $P$ and $S$ (that we suppose not antipodal!).
The astronomical object $X$ under simultaneous study of $P$ and $S$ is 
also characterized by a cosmic latitude $\ell(X)$. It is obvious that 
$\ell(P)=\frac{\pi}{2}$. It is also clear that the (conformal) time difference 
$\delta = \tau_o - \tau$ between observation (by $S$ at time $\tau_o$) and 
emission of light (by $X$ at time $\tau$) is nothing else than a 
measure of the arc of geodesic defined by the two points $S$ and $X$ 
on the unit three-sphere. We have therefore a ``triangle'' $PXS$ whose 
``edges'' are arcs of great circles and the lengths of these three arcs 
are $b=\frac{\pi}{2} - \ell(S)$ (from $P$ to $S$), $a=\frac{\pi}{2} - 
\ell(X)$ (from $P$ to $X$) and $c=\delta$ (from $S$ to $X$).
The last piece of information that we need is a measure $\alpha$ of 
the angle\footnote{This symbol $\alpha$ has of course nothing to do
with the one that was previously used to denote the radiation 
parameter}, as seen from Sun ($S$) between the direction of $P$ and
the direction of $X$. It is a priori clear (draw a triangle !) that
there is a relation between $\ell(S)$, $\ell(X)$, $\delta$ and 
$\alpha$, a relation that generalizes the well known 
formula $a^2=b^2+c^2 - 2 bc \cos \alpha$ valid for an arbitrary 
triangle in euclidean space. Here is the formula that we need:
\begin{equation}
\sin \ell(X) = \cos \delta \sin \ell(S) - \sin \delta \cos \alpha \cos \ell(S)
\label{eq:SphericalTriangle}
\end{equation}
The proof of this formula is given in Appendix ~2 and uses the fact 
the three-sphere $S^3$ carries a group structure: it can be identified with $SU(2)$ or with 
the unit sphere in the non-commutative field of quaternions $\HH$.
The uninterested reader may take the above 
formula for granted  but the technique 
used in our proof is of independent interest and may be used by the 
reader to solve problems of similar nature.

The formula \ref{eq:SphericalTriangle} gives in particular  the 
(conformal) time difference $\delta$ between $S$ and 
$X$, when all the points of our spatial universe are characterized in 
terms of cosmic latitude with respect to an arbitrary reference point $P$ (the pole).
The equation giving $\delta$ should be solved numerically in general, 
but notice that the equation simplifies, and give rises to an analytic
expression, when we choose $X$ on the equatorial two-sphere defined 
by $P$. In that particular case, $\ell(X) = 0$ and
\begin{equation}
\delta  = \arctan(\frac{\tan \ell(S)}{\cos (\alpha)})
\label{eq:SphericalTriangleCasEquatorial}
\end{equation}
 
Coming back to our problem of comparing redshifts, we find that the
redshift of the object $X$, as observed by $P$ at conformal time 
$\tau_o$, is 
$$
Z = C [{\cal P}(\tau_o - (\pi/2 - \ell(X)) ) + \frac{1}{12}] - 1
$$
with $C$ still given by equation \ref{eq:RedshiftConstant}, 
whereas the redshift of the same object $X$, as observed by $S$ (Sun) at the 
same time $\tau_o$ is given by equation \ref{eq:RedshiftFromS},  with
$\delta$ determined by equation \ref{eq:SphericalTriangle}; as before,
$\ell(X)$ and $\ell(S)$ refer to the cosmic latitude of $X$ and $S$ 
with respect to the reference point $P$ (pole) and $ \alpha$ is the 
angle between the sighting directions $SP$ and $SX$.

In the particular case of an astronomical object $X$ belonging to the two-sphere
which is equatorial with respect to the reference point
$P$, the formulae can be simplified:
The redshift of $X$ as, recorded by $P$ is then 
\begin{equation}
Z =\frac{ [{\cal P}(\tau_o - \pi/2) + \frac{1}{12}]}{[{\cal P}(\tau_o) + 
\frac{1}{12}]} - 1
\label{eq:RedshiftFromPCasEquatorial}
\end{equation}
whereas, as recorded by $S$, it is 
\begin{equation}
z = \frac{ [{\cal P}(\tau_o  - \arctan(\frac{\tan
\ell(S)}{\cos (\alpha)})) + \frac{1}{12}]} {[{\cal
P}(\tau_o) + \frac{1}{12}]} - 1
\label{eq:RedshiftFromSCasEquatorial}
\end{equation}

The redshift of equatorial objects $X$, as seen from $P$, can be numerically computed from equation 
\ref{eq:RedshiftFromPCasEquatorial}, and it is a direction-independent quantity.
The redshift of the same equatorial objects, as seen from $S$ 
is clearly direction dependent.
The existence of a direction-independent significative  gap in the distribution of 
quasars, as seen from 
a particular point $P$ of the Universe, around the value $Z$ of the 
redshift, would certainly be an example of a (remarkable) large scale structure,
but such a gap would become direction dependent as seen from another 
point $S$ (the Sun); the previously given formulae would then be 
necessary to perform the necessary change of redshift charts allowing 
one to recognize the existence of these features.
\footnote{existence of such a gap was investigated in references \cite{Souriau}}

Notice however that, in this particular example ($X$ equatorial with respect 
to $P$) we have to suppose $\tau_o > \pi/2$ since, in the opposite case, 
the light coming from $X$ cannot be recorded by $P$. Such an hypothesis 
is not necessarily satisfied and actually  is {\sl not } 
satisfied when we use the values 
given in our numerical example of section 4.3.2 since we found, in 
that case, $\tau_{o}= 1.369< \pi/2$. In such a situation, if we want to
compare measurements between $S$ and $P$ made at the same time, we have 
to look at an object $X$ which is not equatorial with respect to $P$ 
but is such that its light can reach both $S$ 
and $P$ at time $\tau_{o}$; the first condition (observability from 
$P$) gives $\pi/2 - \ell(X) < \tau_{o}$ \ie $\ell(X) > \pi/2 - \tau_{o}$ whereas the second one 
(observability from $S$) reads $\delta < \tau_{o}$, but $\delta$ is 
still given by \ref{eq:SphericalTriangle}, so that this condition 
translates as a condition on the angle $\alpha$ between the sighting 
directions of $X$ and of $P$ as seen from $S$ : this angle should be 
small enough with a maximal value obtained by replacing $\delta$ 
by $\tau_{o}$ in equation \ref{eq:SphericalTriangle} and solving for 
$\alpha$.

Another possibility to illustrate the above formulae is
to compare the  redshifts of $X$ measured
at time $\tau_{o}$ from 
two positions $S_{1}$ and $S_{2}$ 
that should have small enough cosmic latitudes since
the two conditions on $\delta_{1}$ and $\delta_{2}$ (the 
geodesic distances between $X$ and these two positions) are
$\delta_{1}< \tau_{o}$ and $\delta_{2}< \tau_{o}$.
To simplify the calculation we suppose that $S_{1}$ and $S_{2}$ belong 
to the same meridian (going through $P$) of our spatial hypersphere, and call
$\alpha_{1}$ (resp. $\alpha_{2}$) the angle made between the sighting 
directions of $X$ and 
of $P$, as seen from $S_{1}$ (resp. from $S_{2}$).
Using the technique explained in Appendix 2 (write $\overline{S_{2}}X = 
\overline{S_{2}}S_{1}\overline{S_{1}}X$), the reader will have no 
difficulty to prove the following formula that generalizes equation
\ref{eq:SphericalTriangle}:
\begin{equation}
\delta_{2} = \arccos(\cos(\ell(S_{2}) - \ell(S_{1})) \cos \delta_{1} 
- \sin(\ell(S_{2}) - \ell(S_{1})) \sin \delta_{1} \cos \alpha_{1}
\label{eq:GeneralSphericalTriangle}
\end{equation}
To simplify even further the calculation, we take $X$ equatorial with respect to the reference 
point $P$; therefore the previous inequalities on $\delta_{1}$ and $\delta_{2}$, using equation 
\ref{eq:SphericalTriangleCasEquatorial}, give, in turn, the following two 
conditions on the cosmic latitudes of the two observers:
$\tan \ell(S_{1}) < \tan \tau_{o} \cos \alpha_{1}$ and 
$\tan \ell(S_{2}) < \tan \tau_{o} \cos \alpha_{2}$.
The redshift $z_{1}$ of $X$ as measured by $S_{1}$ is still given by equation  \ref{eq:RedshiftFromS}
but we can use the particular equation 
\ref{eq:SphericalTriangleCasEquatorial}, valid for equatorial 
objects, therefore 
$z_{1} = C [{\cal P}(\tau_o - \delta_{1}) + \frac{1}{12}] - 1$ 
with $\delta_{1}  = \arctan(\frac{\tan \ell(S_{1})}{\cos (\alpha_{1})})$
whereas
$z_{2} = C [{\cal P}(\tau_o - \delta_{2}) + \frac{1}{12}] - 1$
and $\delta_{2}$ given by equation \ref{eq:GeneralSphericalTriangle}.

In order to illustrate these results in the study of large 
scale structures, we choose, for  the cosmological 
parameters,
the particular values given in section 4.3.2; furthermore we fix the values
of $\ell(S_{1})$  (take $\pi/6$) and of $\alpha_{1}$  (take $\pi/4$).
Then $\delta_{1}$, is given by \ref{eq:SphericalTriangleCasEquatorial},
 and the redshift of $X$, as measured by $S_{1}$  is  $z_{1}=3.42$.
The redshift $z_{2}$ of $X$, as measured by $S_{2}$ is given by the above 
formula, but since everything else is fixed, it becomes a function of
the cosmic latitude $\ell(S_{2})$ only. For definitness we choose
$P$, $S_1$ and $S_2$ in the same hemisphere.
The following figure (\ref{fig:redS1S1.eps}) gives
$z_{2}$ as a function of $\ell(S_{2})$ in the range $[{\pi/6 - 
\pi/12},{\pi/6 + \pi/12}]$; of course this curve intersects the horizontal 
line $z_{1}=3.42$ when $\ell(S_{2}) = \ell(S_{1}) = \pi/6$.

\begin{figure}
\includegraphics{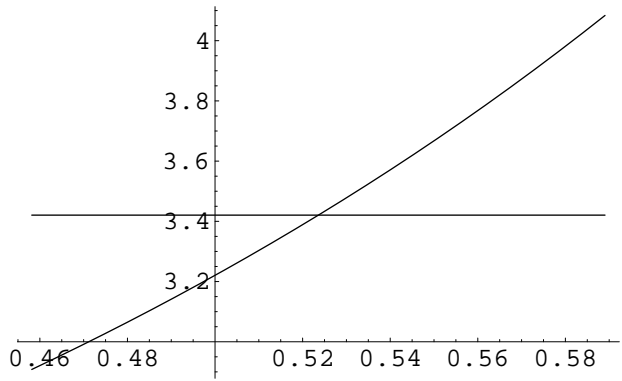}
\caption{Dependence of the redshift function on the cosmic latitude 
of an observer}
\label{fig:redS1S1.eps}
\end{figure}

\subsubsection{Comparison of Measurements Relative to a Single Event}

We now examine the situation where both observers (take $P$ and $S$) are interested in the same 
cosmological event (a star belonging to a distant galaxy turning to a 
supernova, for example) and record the redshifted 
light coming from this event when the light reaches them (both 
measurements are not usually performed at the same time).
Here we do not have to assume that $\tau_{o} > \pi/2$, even if $X$ is 
equatorial with respect to $P$ since the light 
emitted by $X$ will reach $P$ anyway, in some future. Keeping the 
same notations as before, the 
cosmological event of interest took place at conformal time 
$\tau_{o} - \delta$, and the light will reach $P$ at time $\tau_{o} - 
\delta + \pi/2 - \ell(X)$. The redshift measured by $S$ is, as 
before, given by equation \ref{eq:RedshiftFromS} \ie
$$
z = \frac{[{\cal P}(\tau_o - \delta) + \frac{1}{12}]}{ [{\cal P}(\tau_o) + \frac{1}{12}]} -1
$$
but the redshift measured by $P$ is 
$$
Z = \frac{[{\cal P}(\tau_o - \delta) + \frac{1}{12}]}{ [{\cal P}(\tau_o-\delta + \pi/2 - \ell(X)) + \frac{1}{12}]} -1
$$

\appendix
\section{The Case With Radiation ($\alpha \neq 0$). Analytic Solution.}
\paragraph{General features}
When $\alpha \neq 0$, the two poles of $T(\tau)$ are distinct (and
each of them is of first order). The three roots $e_1,e_2,e_3$ of the equation
$P(y) = 0$ (\ref{WeierstrassDiffEq}) add up to zero. They are given  by $- {A+B \over 2} \pm i
{\sqrt{3} \over 2} (A-B)$
and $A+B$, where
\begin{eqnarray}
\label{e123}
A &=& 1/2 [g_{3} + \sqrt{- \Delta/27}]^{1/3} \\
B &=& 1/2 [g_{3} - \sqrt{- \Delta/27}]^{1/3}
\end{eqnarray}
and where $\Delta$ is the discriminant
$$
\Delta = g_{2}^{3} - 27 g_{3}^{2} = 3^{-3} \alpha^{3} \lambda (\lambda -
\lambda_{+}) (\lambda - \lambda_{-})
$$
with
$$
\lambda_{\pm} = {1\over 32 \alpha^{3}} (24 k^{2}\alpha^{2} + 12 k
\alpha + 1 \pm (8 k \alpha + 1)^{3/2})
$$
As $\alpha \rightarrow 0$, we have $$\Delta \rightarrow 2^{-4} 3^{-3}
\lambda (1 - \lambda)$$
indeed, $\lambda_{+}\rightarrow \infty$
and $\lambda_{-} \rightarrow 1$ and, to first order in $\alpha$, we have
\begin{eqnarray*}
\lambda_{+} &\simeq &{1 \over 16 \alpha^{3}} (1 + 12 k \alpha) \\
\lambda_{-} &\simeq & k (1-3\alpha k)
\end{eqnarray*}
If $\Delta = 0$ -- something that is nowadays ruled-out
experimentally  but that used to be called the ``physical case'' ! --
two zeroes of $P(\tau)$ coincide (then also, two
zeroes of $Q(\tau)$); in this situation one of the two periods becomes
infinite and the elliptic function degenerates to trigonometric or
hyperbolic functions. The analytic study, in that situation, is well known.

\paragraph{We know assume that $\alpha \neq
0$ and $\lambda_{-} < \lambda < \lambda_{+}$}
This is the situation that seems to be in agreement with the recent
experiments (see the discussion given previously).
For such values of $\lambda$, the polynomial $Q(T)$ has two
negative roots $T_{a}<T_{b}<0$ (see fig \ref{potential.eps}). Here
$g_{3}<0$ and the discriminant $\Delta$ is (strictly) negative, as well. The
three roots of the equation $P(y) = 0$ can be written
\begin{eqnarray*}
e_{1} = a-ib \\
e_{2} = -2 a \\
e_{3} = a + ib
\end{eqnarray*}
with $a>0$ and $b>0$.

In the case $\alpha \neq 0$, it is of course still possible to
express $T$ in terms of $y$ (using the fractionally linear
transformation \ref{eq:transformationT}, with $y = {\cal
P}(\tau;g_{2},g_{3})$, the Weierstrass ${\cal P}$-function. This is
actually not very convenient and
it is much better to use either the Weiertrass $\zeta$-function
$\zeta (\tau)$ or the Weierstrass $\sigma$-function $\sigma(\tau)$.

\begin{figure}
\includegraphics{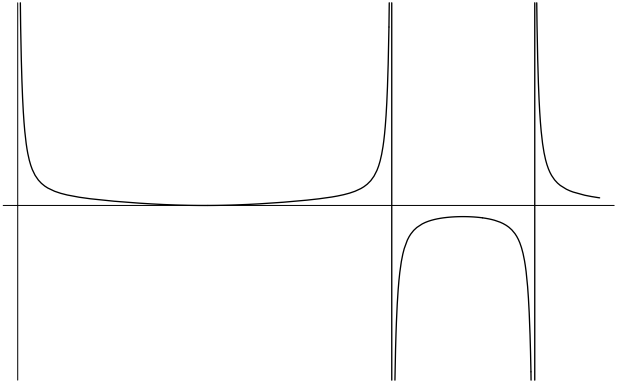}
\caption{Typical evolution of the reduced temperature (case with radiation)}
\label{fig:temperaturewithrad.eps}
\end{figure}

The function $\zeta (\tau)$ -- which is an odd function of $\tau$ is not
elliptic but it is defined by
$$
{d\zeta \over d \tau} = - {\cal P}(\tau)
$$ and the requirement that $1/\tau - \zeta(\tau)$ vanishes at $\tau =0$.
Although $\zeta$ is not elliptic, the difference $\zeta(\tau - u) - \zeta(\tau -
v)$ is elliptic of order 2, with poles at $u$ and $v$, when $u \neq
v$ are two arbitrary complex numbers. In terms of
$\zeta$, the quantities $R(\tau)$ and $T(\tau)$ read immediately:
\begin{equation}
\sqrt{{\Lambda \over 3}} R(\tau) = \sqrt{{\lambda \over 3}} {1 \over
T(\tau)} = \zeta(\tau - \tau_{g}) + \zeta(\tau_{g}) - \zeta(\tau-\tau_{f})
- \zeta(\tau_{f})
\label{eq:TfromWeierstrassZeta}
\end{equation}
where the two zeros of $T(\tau)$,
$0<\tau_{f}<\tau_{g}<2\omega_{r}$ are given by (use figure
\ref{potential.eps}) :
$$
\tau_{f} = \int_{0}^{\infty} {dT \over \sqrt{Q(T)}}
$$
$$
\tau_{g} = \tau_{f} + 2  \int_{T_{b}}^{0} {dT \over \sqrt{Q(T)}}
$$
The quantities $T_{a}$ and $T_{b}$ are defined on the graph given on 
figure \ref{potential.eps}; they are numerically determined by solving 
the equation $V_{\alpha k}(T) = \frac{\lambda}{3}$ (for instance use 
the function  FindRoot of Mathematica). The curve
$T(\tau)$, in the closed case, has the shape given by figure \ref{fig:temperaturewithrad.eps}
 in full agreement with the qualitative discussion given in  section 4;
of course, only the first branch of this curve, from the big bang ($\tau = 0$) to
the first zero $\tau_f$ of $T(\tau)$ (the end of conformal time) is physically relevant.
It is instructive to compare this function with
with the $\alpha = 0$ case (no radiation), given in figure
\ref{TgraphClosed.eps}. In both cases, we see that
the function $T(\tau)$ is negative between $\tau_f$ and $\tau_g$ (its next zero),
but here, it 
becomes infinite again for a value $\tau_f + \tau_g$ of conformal time
which is strictly smaller than $2 \omega_{r}$; 
after this singularity, we have a negative branch (a novel feature of the case with radiation)
in the interval $]\tau_f + \tau_g, 2\omega_r[$; the period, on the real axis is, as usual, 
denoted by $2\omega_r$. One shows that
$
2\omega_{r} = \tau_{f} + \tau_{g} + \tau_{c}
$
where
$$
\tau_{c} = 2 \int_{-\infty}^{T_{a}} {dT \over \sqrt{Q(T)}}
$$

Such  plots were already given in \cite{CoqueGros} where
Weierstrass functions had been numerically calculated from the algorithms
obtained in the same reference. 

The same reduced temperature can be expressed in terms of the Weierstrass function $\sigma(\tau)$.
The even function $\sigma (\tau)$ is not elliptic either and  is defined by
$$
{1 \over \sigma(\tau)} {d \sigma\over d \tau} =\zeta (\tau)
$$ and the requirement that $\sigma(\tau)$ should be an entire function
vanishing at $\tau =0$.
Although not elliptic, the quotient
${\sigma(\tau - u_{1}) \sigma(\tau - u_{2}) \over \sigma(\tau - u_{3})
\sigma(\tau - u_{4})}$ is elliptic of order 2,
  with poles at $u_{3}$, $u_{4}$ and zeroes at $u_{1}$, $u_{2}$,
  whenever $u_{1}, u_{2}, u_{3}, u_{4}$ are complex numbers such that
  $u_{1}+ u_{2}= u_{3}+ u_{4}$ . In terms of
$\sigma$, the quantity $T(\tau)$ reads immediately:
$$
T(\tau) = c .  {\sigma(\tau - \tau_{f}) \sigma(\tau - \tau_{g}) \over
\sigma(\tau) \sigma(\tau -  \tau_{f}- \tau_{g})}
$$
where the constant $c$ is given by
$$
c = T_{b} {\sigma^{2}({\tau_{g}+\tau_{f} \over 2}) \over
\sigma^{2}({\tau_{g}-\tau_{f} \over 2})}
$$

\section{Geometrical Relations in a Geodesic Triangle}

The unit three-sphere $S^3$ will be here identified with 
the unit sphere in the non-commutative field of quaternions $\HH 
\simeq \RR^4$. To each $X \in S^3 \subset \RR^4$, $X = (X_0, 
X_{1},X_{2},X_{3})$ we associate the quaternion
\begin{equation}
X = X_{0}+X_{1} i+X_{2} j+X_{3} k
\label{eq:GenericQuaternion}
\end{equation}
We shall use the non commutative multiplication rules $i^2=j^2=k^2=-1$, $ij=-ji=k$, 
$jk=-kj=i$ and $ki=-ik=j$.
The norm square of $X$ is given by $\vert X \vert^2 = {\overline 
X}X = X_{0}^2+X_{1}^2+X_{2}^2+X_{3}^2$ where ${\overline X}$ is the 
quaternionic conjugate of $X$ ( with ${\overline i}=-i,{\overline 
j}=-j, {\overline k}=-k) $.

It is convenient to write $X$ in a form analoguous to the 
representation $X= e^{i\delta}=\cos \delta + i \sin \delta$ of a complex number 
of unit norm; in this elementary situation,  the angle $\delta$ is the geodesic distance (arc 
length) between the unit $1$ and the complex $X$, whereas $i$ (unit norm) 
can be considered as the tangent (sighting) direction from $1$ to $X$ along an arc
of geodesic (since $i$ represents the vector $(0,1)$ tangent to the 
unit circle at the point $(1,0)$). In the three dimensional situation, 
the unit $1$ is a particular point of the three-sphere $S^3$ and
we shall write, in the same way 
\begin{equation}
X = e^{{\hat X}\delta} = \cos \delta + {\hat X} \sin \delta
\label{eq:ExpNotationForQuaternion}
\end{equation}
where $\delta$ is the geodesic distance (arc 
length) between the unit $1$ and the quaternion $X$, while ${\hat 
X}$ is a quaternion of square $-1$ and unit norm representing the three dimensional vector 
tangent to $S^3$ at the point $1$ in the direction $Y$ (sighting 
direction).
More precisely, let  ${\hat X}_{\ell}$, $\ell = 1,2,3$, be this three 
dimensional unit vector (${\hat X}_{1}^2 + {\hat X}_{2}^2 + {\hat 
X}_{3}^2 = 1$) and call
$${\hat X} \doteq {\hat X}_{1} i + {\hat X}_{2} k + {\hat X}_{3} k$$
Then ${\hat X}{\hat X}=-1$ and ${\overline {\hat X}}{\hat X} = +1$.
Comparing equations \ref{eq:GenericQuaternion} and
\ref{eq:ExpNotationForQuaternion} gives 
\begin{equation}
{\hat X}_{\ell} = \frac{X_{\ell}}{\sin \delta}
\label{eq:sightingdir}
\end{equation}
and
\begin{equation}
\cos \delta = X_{0}
\label{eq:geodesiclength}
\end{equation}

Take ${\overrightarrow u}$ and ${\overrightarrow v}$ two three-dimensional 
vectors (not necessarily of unit norm), call $u = u_{1} i + u_{2} 
j + u_{3} k$, $v=v_{1}i + v_{2} j + v_{3}k$ the corresponding 
quaternions and ${ {\hat u}}$, ${ {\hat v}}$ the 
corresponding  normalized  unit vectors.
An elementary calculation, using the previous multiplicative 
rules, leads to
$uv = - \cos \alpha + {\overrightarrow u} \times {\overrightarrow v}$,
but $\vert {\overrightarrow u} \times {\overrightarrow v} \vert = \vert {\overrightarrow u}
\vert \, \vert {\overrightarrow v} \vert \, \sin \alpha$,
so that
$$u \, v = \vert {\overrightarrow u} \vert \, \vert {\overrightarrow v} 
\vert (- \cos \alpha +  {\widehat {{\overrightarrow u} \times {\overrightarrow 
v}}} \sin \alpha ) $$
where $\alpha$ is the angle between vectors ${\overrightarrow { u}}$, ${\overrightarrow 
{ v}}$, and ${\widehat {{\overrightarrow u} \times {\overrightarrow 
v}}}$ is the normalized vector product of these two 
three-dimensional vectors. For normalized vectors, 
we have in particular
$(uv)_{0} = - \cos \alpha$, or better,
\begin{equation}
({\overline u} \, v)_{0} = + \cos \alpha
\label{eq:AngleEquation}
\end{equation}

Now, let $P$ (a reference point ``Pole''), $S$ (Sun) and $X$ (a quasar) 
three points of $S^3\subset \HH \simeq \RR^4$ that 
we represent by quaternions also denoted by $P$, $S$ and $Y$. 
We call $\ell(S)$, resp. $\ell(X)$, the latitude of $S$ (resp. of $X$) 
with respect to $P$, counted positively whenever $P$ and $S$ (resp. 
$X$) are in the same hemisphere, and counted negatively otherwise.

Let $Y \doteq {\overline S}X$. From equations \ref{eq:sightingdir} and \ref{eq:geodesiclength}, we 
find that the geodesic distance $\delta$ between $X$ and $S$ (or 
between $1$ and $Y$) is such that $\cos \delta = Y_{0}$ so,
\begin{equation}
\delta = arc cos ({\overline S}X)_{0}
\label{eq:GeodesicResult}
\end{equation}
and that the sighting direction from $S$ to $X$ is given by a unit 
vector ${{\hat Y}}$ with three components 
\begin{equation}{\hat Y}_{\ell} = \frac{{\overline S}X_{\ell}} {\sin \delta}
\label{eq:VectorResult}
\end{equation}
We have
$$
Y = {\overline S}X = \cos \delta + {\hat Y} \sin 
\delta  =\cos \delta + {\widehat {{\overline S}X}} \sin \delta 
$$

Notice that, in the same way,
$$
{\overline P}S  = \cos(\frac{\pi}{2} - \ell(S)) +  \sin(\frac{\pi}{2} - 
\ell(S)) {\widehat{ {\overline P}S}} = \sin(\ell(S)) + \cos(\ell(S)) {\widehat{ {\overline P}S}}
$$
and
$$
{\overline P}X = \sin(\ell(X)) + \cos(\ell(X)) {\widehat{ {\overline P}X}}
$$

Using the fact that $S{\overline S}=1$, we remark that 
$${\overline P}X = {\overline P}S {\overline S}X$$

This last equality provides the clue relating the three sides of our 
geodesic triangle. It implies in particular
$({\overline P}X)_{0} = ({\overline P}S {\overline S}X)_{0}$. The left 
hand side is given by
$({\overline P}X)_{0} = \sin(\ell(X))$ and the right hand side comes from
\begin{eqnarray*}
    ({\overline P}S )({\overline S}X) &=& (\sin(\ell(S)) + \cos(\ell(S)) 
    {\widehat{ {\overline P}S}})(\cos \delta + \sin \delta \, {\widehat {{\overline S}X}}) \\
    {} &=& \sin \ell(S) \, \cos \delta + \cos \ell(S) \, \sin \delta {\widehat{ 
    {\overline P}S}}\, {\widehat {{\overline S}X}} + \\
    {} & {} & \sin \ell(S) \, \sin \delta  \, {\widehat {{\overline S}X}} + \cos \ell(S) \, 
    \cos \delta {\widehat{ {\overline P}S}} \\
     (({\overline P}S )({\overline S}X))_{0} &=& \sin \ell(S) \, \cos \delta + 
     \cos \ell(S) \, \sin \delta ({\widehat{ 
    {\overline P}S}}{\widehat{ {\overline S}X}})_{0}
\end{eqnarray*}
By equation \ref{eq:AngleEquation}, we know that the cosine of the angle $\alpha$ 
between vectors $\overrightarrow { SP}$ and 
$\overrightarrow { SX}$, respectively represented by the quaternions 
${ {\overline S}P}$ and ${{\overline S}X}$, is given by the $0$-component of the 
quaternionic product ${\overline {\widehat {({\overline S}P)}}}({\widehat {{\overline S}X}}) = 
(\widehat{{\overline P}S} )(\widehat{{\overline S}X}) $. 

To conclude, one finds the following formula relating the lengths of 
the three sides of the geodesic triangle $SXP$, together with the angle 
$\alpha$:
$$
\sin(\ell(X)) = \sin \ell(S) \, \cos \delta - 
     \cos \ell(S) \, \sin \delta
     \, \cos \alpha
$$



\end{document}